\pdfoutput=1

\documentclass[journal,onecolumn,draftclsnofoot]{IEEEtran}
\usepackage{cite}

\usepackage[para]{threeparttable}
\usepackage{footnote}
\makesavenoteenv{tabular}

\ifCLASSINFOpdf
   \usepackage[pdftex]{graphicx}
   \graphicspath{{../figures}}
   \DeclareGraphicsExtensions{.pdf,.jpeg,.png}
\else
   \usepackage[dvips]{graphicx}
   \graphicspath{{../figures}}
   \DeclareGraphicsExtensions{.eps}
\fi
\usepackage{amsmath}
\usepackage{amssymb,mathtools}
\usepackage{romannum}
\usepackage[shortlabels]{enumitem}
\usepackage{xcolor}

\newtheorem{theorem}{Theorem}

\newtheorem{lemma}[theorem]{Lemma}

\usepackage{algorithm}
\usepackage{algorithmic}
\usepackage{eqparbox,array}

\ifCLASSOPTIONcompsoc
  \usepackage[caption=false,font=normalsize,labelfont=sf,textfont=sf]{subfig}
\else
  \usepackage[caption=false,font=footnotesize]{subfig}
\fi
\begin{document}
%
\title{Efficient Solvers for Wyner Common Information with Application to Multi-Modal Clustering} 
%
%
%

\author{Teng-Hui~Huang,~\IEEEmembership{Member,~IEEE,}
        and~Hesham~El Gamal,~\IEEEmembership{Fellow,~IEEE}
\thanks{An earlier version of this paper was presented in part at the 2023 IEEE International Symposium on Information Theory (ISIT) [DOI:10.1109/ISIT54713.2023.10206810]. (Corresponding author: Teng-Hui Huang.)}
\thanks{T-H. Huang and H. El Gamal are with the School
of Electrical and Computer Engineering, Univeristy of Sydney, NSW,
Australia.
E-mail: \{tenghui.huang, hesham.elgamal\}@sydney.edu.au}
}

%
%

\markboth{Under review for IEEE Transaction on Information Theory}%
{Huang \MakeLowercase{\textit{et al.}}: Efficient Solvers for Wyner Common Information with Application to Multi-Modal Clustering}
%



\maketitle
\pagenumbering{arabic} 
\begin{abstract}
We propose two novel extensions of the Wyner common information optimization problem. Each relaxes one fundamental constraints in Wyner's formulation. The \textit{Variational Wyner Common Information} relaxes the matching constraint to the known distribution while imposing conditional independence to the feasible solution set. We derive a tight surrogate upper bound of the obtained unconstrained Lagrangian via the theory of variational inference, which can be minimized efficiently. Our solver caters to problems where conditional independence holds with significantly reduced computation complexity; 
On the other hand, the \textit{Bipartite Wyner Common Information} relaxes the conditional independence constraint whereas the matching condition is enforced on the feasible set. By leveraging the difference-of-convex structure of the formulated optimization problem, we show that our solver is resilient to conditional dependent sources. Both solvers are provably convergent (local stationary points), and empirically, they obtain more accurate solutions to Wyner's formulation with substantially less runtime. 
Moreover, them can be extended to unknown distribution settings by parameterizing the common randomness as a member of the exponential family of distributions. Our approaches apply to multi-modal clustering problems, where multiple modalities of observations come from the same cluster. Empirically, our solvers outperform the state-of-the-art multi-modal clustering algorithms with significantly improved performance.
\end{abstract}


%
\IEEEpeerreviewmaketitle

\section{Introduction}
%
%
%
%
\IEEEPARstart{I}{n} the seminal work by Wyner~\cite{wyner2sources}, the information shared among two correlated sources has been characterized. The operational meaning is that two correlated sources can be expressed into a triplet of random variables, consisting of a common random variable, and two statistically independent ones~\cite{xulossy16}. The Wyner common information is formulated as the following problem:
\begin{IEEEeqnarray}{rCl}
\underset{X_1\rightarrow Z\rightarrow X_2}{\min}\,& I(X_1,X_2;Z),\label{eq:wyner_2source}
\end{IEEEeqnarray}
where $Z$ denotes the common randomness of two correlated sources $X_1,X_2$; the joint distribution $P(X_1,X_2)$ is known; the variables to optimize with are the conditional probability $P(Z|X_1,X_2)\in\Omega$ belonging to a compound simplex $\Omega$, formed by cascading each probability simplex of the realizations of $x_1\in\mathcal{X}_1,x_2\in\mathcal{X}_2$. The optimization problem \eqref{eq:wyner_2source} is non-convex and hence difficult to solve~\cite{kumar2014exact}. Yet, the common randomness for two correlated sources is characterized in Wyner's seminal work for discrete settings. A later effort has extended the characterization to continuous Gaussian sources~\cite{xulossy16} and discrete doubly symmetric binary sources~\cite{yu2023dsbs}. Beyond these special cases, the characterization for general settings is less explored until recently. In~\cite{hanna2023cid}, the common information dimension is introduced to characterize the dimensions required for distributed simulation of the common information of joint Gaussian sources. The equivalence between the Hirschfeld-Gebelein-R{\'e}nyi maximal correlation and the local decomposition of the Wyner common information is characterized~\cite{huang2020localWyner}. Notably,~\cite{huang2020localWyner} identified that Wyner common information can be applied to multi-modal learning, where sources of observations share a common target. In~\cite{huang2020localWyner} the empirical covariance matrices of multi-modal data samples are computed for applying the algorithm proposed therein. Complement to these recent theoretic results, we focus on efficient computation algorithms solving (extensions of) Wyner common information problems and provide a new principled framework for unsupervised multi-modal learning. Our motivation is more than the applications. Additionally, we propose that the Wyner common information is ideally suited for capturing the ``shared bits'' between multiple modalities of data. Based on this insight, we develop computation efficient solvers with improved performance over the state-of-the-art methods in a range of multi-modal datasets of varying formats of data and different number of data modalities.

\paragraph*{Notations}
The upper-case letter $Z$ denotes random variables and lower case $z\in\mathcal{Z}$ for realizations (the calligraphic $\mathcal{Z}$ for alphabets). The cardinality of a random variable is denoted as $|\mathcal{Z}|$. Bold-face lower-case letter $\boldsymbol{x}$ and upper-case $\boldsymbol{X}$ refer to a vector and a matrix respectively. The letter $V$ refers to the number of sources, and the index set is defined as $[V]:=\{1,\cdots,V\}$; $\Pi_V:=\{S|\forall S\subset[V],S\cup S^c=[V],|S|\leq|S^c|\}$ denotes all possible bipartite of the index set $[V]$. For observations of all sources, we define $X^V:=(X_1,\cdots,X_V)$, where for each element, the subscript denotes the source index. With a slight abuse of notation, if $S$ is a set, we denote $X_S=(X_{i_1},\cdots,X_{i_{|S|}})$ where $i_j\in S,\forall j\in[|S|]$. $X_1\perp X_2$ represents statistical independence of $X_1,X_2$. $P(X^V)$ denotes the joint distribution of all sources of observations whereas the subscript $\theta$ in $P_\theta(X^V)$ indicates that the distribution is parameterized. The parameter space is often denoted as $\Theta$. $\Omega$ denotes a feasible set which might come with a subscript to indicate the dependency. $D_{KL}[\mu\parallel \nu]$ denotes the Kullback-Leibler divergence between two proper probability measures~\cite{cover1999elements}.
And $\mathcal{O}(\cdot)$ is the standard Big-O notation~\cite{cormen2022introduction}.

\subsection{Extensions of Wyner Common Information}
The extensions of Wyner's original formulation in literature can be divided into three categories. The first line of work changes the optimization objective functions. In~\cite{kumar2014exact}, the \textit{Exact common information} minimizes the Shannon entropy of the common information variable, i.e., $H(Z)$, instead of the mutual information $I(Z;X_1,X_2)$ in~\eqref{eq:wyner_2source} for minimum-length description of $Z$. The connections when the exact variant and Wyner's formulation are equal is later characterized in~\cite{Vellambi2018connect}; In a recent work \cite{Yu2020renyiwyner}, the authors further extend the exact common information to $\infty$-Renyi entropy with characterization provided therein.

Another body of work, aligned with our focus, relaxes the constraints of Wyner's formulation. The relaxed problems are computationally less complex, allowing the use of simpler solvers whose solutions are then mapped to the feasible solution set of Wyner's formulation through projection. In~\cite{sula2021gray}, the \textit{Relaxed Wyner common information} allows the conditional mutual information to be non-zero $I(X_1;X_2|Z)\leq\varepsilon$, where $\varepsilon>0$ is a tolerance threshold. The characterization of the relaxed common information for correlated Gaussian sources is shown in~\cite{sula2021gray}. The method is later extended to $V$ discrete sources in a follow-up work~\cite{sula2021common}, where the constraint is extended as $\sum_{i=1}^VH(X_i|Z)-H(X^V|Z)\leq \varepsilon$. Contrary to this method that does not fully take the correlations between sources into considerations, in our earlier work~\cite{huang2023efficient}, we propose a natural extension leveraging the conditional independence conditions~\cite{ugur2020condindp}: $X_S\rightarrow Z\rightarrow X_{S^c},\forall S\in\Pi_V$ (Equivalently, the set of constraints are: $I(X_S;X_{S^c}|Z)=0$). Then the relaxation naturally follows by: $I(X_S;X_{S^c}|Z)\leq\varepsilon_S$ where $\varepsilon_S\geq 0$ are control thresholds. We call the latter approach the \textit{Bipartite Wyner Common Information} in this work.

The last line of works changes the feasible solution set, followed by a projection onto a solution of Wyner's formulation. In our earlier work~\cite{huang2023efficient}, the solution set of the proposed \textit{Variational Wyner Common Information} satisfies the conditional independence, i.e., $P_\theta(X^V|Z):=\prod_{i=1}^VP_\theta(X_i|Z)$. Following this, $\theta$ is then projected to the feasible solution set in Wyner's formulation:
\begin{IEEEeqnarray}{rCl}
    P_\theta(z|X^V)=\frac{P_\theta(z)\prod_{i=1}^VP_\theta(X_i|z)}{\sum_{z'\in\mathcal{Z}}P_\theta(z')\prod_{j=1}^VP_\theta(X_j|z')}=\frac{P_\theta(z)\prod_{i=1}^VP_\theta(X_i|z)}{P_\theta(X^V)},
\end{IEEEeqnarray}
Note that $P_\theta(X^V)$ here can be different from $P(X^V)$. As a result, an extra matching constraint is imposed on the variational formulation: $D_{KL}[P(X^V)\parallel P_\theta(X^V)]\leq \xi,\xi>0$, where $\xi$ is a control threshold~\cite{huang2023efficient}. The advantage of the variational form is that the growth rate of the number of parameters with respect to the number of sources is linear $\mathcal{O}(V|\mathcal{Z}||\mathcal{X}|)$ in $V$ whereas it is exponential $\mathcal{O}(|\mathcal{Z}||\mathcal{X}|^V)$ for the relaxed variants~\cite{sula2021gray}.

\subsection{Multi-Modal Learning and Wyner Common Information}
The extensions mentioned in the last section all assume knowing the joint distribution $P(X^V)$. As for unknown $P(X^V)$, the problem becomes prohibitively difficult. Yet, it is often possible to have a set of observations of $X^V$, e.g., a dataset, available for data processing. Recently, it is of great interest to extract ``good'' features that are shared between multiple modalities of data for improved performance~\cite{8715409,gao2020survey,ngiam2011multimodal}. Here, multi-modality refers to data of different sources/forms, e.g., images and audio. Notably, the closely related multi-view learning problem can be treated as a special case of multi-modal learning when the different data streams are of different formats but the same modality, e.g., images with different colors~\cite{2019Mvsurvey,sun2013survey,MvCsurvey2018}.
Following the perspective of finding good ``shared bits'', it is then beneficial to capture this common randomness without estimating the unknown and potentially high-dimensional $P(X^V)$, then directly apply the shared information for downstream tasks (e.g., clustering). The motivation aligns with the recent progress in multi-modal learning where the separation of modality-shared and modality-specific features from multi-modal observations has shown empirical success over earlier approaches~\cite{bekkerman2007mmc_earlier,cai2011hifiv_earlier}. Moreover, the insight is especially suitable for unsupervised learning settings as the multi-modal training samples $X^V$ are observed without knowledge of the common source.

In unsupervised extraction of the modality-shared features in unknown joint distribution settings, we focus on recent information theoretic-based approaches. In particular, the generalized evident lower bound (ELBO)~\cite{sutter2021generalized} extends the well-known ELBO to multi-modal observations and imposes factorization conditions for efficiency in finding shared features of modal-specific representations. Several works have adopted this generative direction, but they propose different methods in forming the common features. They can be categorized into two directions: 1) concatenation and 2) alignment. Concatenation methods consist of two steps. First, each single-modal observation is encoded at a modal-specific rate constraint, then the same compressed feature is decoded for all modalities with all distortion criterion satisfied. Second, the common features are formed from cascading each single-modal encoded message. One representative approach in this category is the mixture-of-experts (MoE)~\cite{shi2019variationalMOE,xu2022MFLVC,palumbo2023mmvae+,hwang2021multi}; Alignment methods also imposes a modal-specific rate constraint on the associated single-modal encoders, but all the encoded modal-specific messages are combined into a shared feature vector instead, e.g., a arithmetic or geometric average. Then the shared vector is decoded by each modal-specific decoder subject to separate distortion constraints. The representative approach in this category is the product-of-experts (PoE)~\cite{wu2018multimodalPOE,poe2020feifeiPoE,hwang2021multi}. The categorization is not strict in the sense that the aligned common features can later be cascaded as a mixture for reconstruction or clustering, such as the Mixture-of-Product-of-experts (MoPoE)~\cite{sutter2021generalized}. It is worth noting that as the ELBO is essentially estimating the unknown joint distribution by introducing an auxiliary representation variable, the common feature in the generalized ELBO is a by-product in estimating the multi-modal joint distribution. On the contrary, Wyner common information focuses on characterizing the common randomness shared between multi-modal observations, instead of the estimation of the joint distribution.

\subsection{Contributions}
We propose efficient solvers for two extensions of the Wyner Common information, namely, the \textit{Variational Common Information} (VCI) and \textit{Bipartite Common Information} (BCI) respectively. These solvers apply to both known and unknown joint distribution settings. For known distribution settings, the proposed iterative solver for \textit{VCI} has linear complexity growth rate with respect to the increase of the number of sources. Then for \textit{BCI}, we propose a novel Difference-of-Convex Algorithm (DCA)-based iterative solver with convergence guarantee to a local stationary point. The proposed \textit{VCI} solver caters to scenarios where the common source can be inferred from the realizations of the observations (invertible) whereas our \textit{BCE} solver obtains better solutions in Wyner's formulation in non-invertible cases. Notably, both the proposed solvers obtain better solutions in Wyner's formulation than the state-of-the-art solvers in knowing distribution settings in our evaluation. 

As for unknown joint distribution settings, we focus on unsupervised multi-modal clustering where the observations for training have multiple modalities that come in pairs and share a hidden common cluster variable (without label). We show that both the \textit{VCI} and \textit{BCI} are efficient for multi-modal clustering since we parameterize the multi-modal observations as a mixture of members of the exponential family. The parameters can be estimated from maximum-likelihood principle and are step-size free, resulting in simpler implementation than earlier gradient descent-based methods with significantly faster running time.

Comparing the two solvers, the \textit{VCI} solver achieves lower computation complexity than that for the \textit{BCI} whereas the latter achieves higher accuracy than the former in general cases where the common source cannot be inferred from the observations. Remarkably, identifying the optimal number of clusters without supervision is part of the proposed solvers in contrast to previous approaches. Finally, we evaluate our solvers in a variety of datasets of different modalities and the numbers of modalities. Our solvers either achieve higher accuracy or lower running time as compared to the state-of-the-art baselines.

\section{Problem Formulation}

\subsection{Bipartite Common Information}
One of difficulties in computing the Wyner common information is because of the Markov relations imposed on the feasible solution set of \eqref{eq:wyner_2source}, i.e., $I(X_i;X_j|Z)=0,\forall i\neq j\in[V]$. In \textit{Bipartite} common information, we relax this restriction. For illustration, we start with the two-modality case $V=2$. Given $P(X_1,X_2)$, the goal is to minimize $I(X_1,X_2;Z)$ such that $X_1\rightarrow Z\rightarrow X_2$ over the conditional probability $P(Z|X_1,X_2)$:
\begin{IEEEeqnarray}{rCl}
    \underset{P_\theta(Z|X_1,X_2)}{\min}&&\quad I(X_1,X_2;Z),\nonumber\\
    \text{subject to}&&\quad I(X_1;X_2|Z)\leq \eta_{1},\IEEEeqnarraynumspace\IEEEyesnumber\label{eq:rep_form_2v_case}
\end{IEEEeqnarray}
for some control threshold $\eta_1>0$. Compared to Wyner's formulation \eqref{eq:wyner_2source}, the difference is that now we allow the conditional mutual information to be upper bounded by a control threshold $\eta_1$. Note that~\eqref{eq:rep_form_2v_case} is a special case of the \textit{Relaxed} common information~\cite{sula2021gray} in literature. But our new insight is that we can further express \eqref{eq:rep_form_2v_case} into an equivalent form with two competing sub-objectives:
\begin{IEEEeqnarray}{rCl}
    I(X_1;X_2|Z)=I(X_1,X_2;Z)-I(X_1;Z)-I(X_2;Z)+I(X_1;X_2).\IEEEeqnarraynumspace\IEEEyesnumber\label{eq:rep_2v_key_relation}
\end{IEEEeqnarray}
Substituting \eqref{eq:rep_2v_key_relation} into \eqref{eq:rep_form_2v_case}, the problem is equivalent to:
\begin{IEEEeqnarray}{rCl}
    \underset{P_\theta(Z|X^V)}{\min}&&\quad I(X_1,X_2;Z),\nonumber\\
    \text{subject to}&&\quad -I(X_1;Z)-I(X_2;Z)\leq \eta_1',\IEEEeqnarraynumspace\IEEEyesnumber\label{eq:rep_compete_2v_case}
\end{IEEEeqnarray}
for some control threshold $\eta_1'>0$. It is straightforward to see that when $P(X_1,X_2)$ is known, then $I(X_1,X_2;Z)$, $I(X_2;Z)$ and $I(X_1;Z)$ are all convex functions of $P_\theta(Z|X_1,X_2)$. Therefore, the overall goal of \eqref{eq:rep_compete_2v_case} can be viewed as minimizing $I(X_1,X_2;Z)$ and maximizing $\sum_{i=1}^2I(X_i;Z)$ simultaneously. 

The \textit{Bipartite} form digresses from the relaxed common information for $V\geq 3$. We illustrate this with a $V=3$ example. Define $\Pi_3:=\{\{1\},\{2\},\{3\}\}$ and the unconstrained Lagrangian is given by:
\begin{IEEEeqnarray}{rCl}
    \mathcal{L}_3:=I(Z;X_1,X_2,X_3)&-&\kappa_{\{1\}}\left[I(Z;X_1)+I(Z;X_1,X_2)\right]\nonumber\\
    &-&\kappa_{\{2\}} \left[I(Z;X_2)+I(Z;X_1,X_3)\right]\nonumber\\
    &-&\kappa_{\{3\}} \left[I(Z;X_3)+I(Z;X_1,X_2)\right],\nonumber
\end{IEEEeqnarray}
where $\{\kappa_S\}_{S\in\Pi_3}$ are a set of non-negative multipliers. Clearly, the \textit{Bipartite} form relaxes the conditional independence conditions by enumerating all the mutual information between the common target $Z$ and each possible bipartite of the information sources as constraints instead.

Then we generalize the problem to an arbitrary number of sources. This can be achieved through partitioning the index set $[V]$ into bipartite. Consider a subset set $S\subset[V]$, the generalized \textit{Bipartite Common Information} optimization problem is given by:
\begin{IEEEeqnarray}{rCl}
    \underset{P(Z|X^V)}{\min}&&\quad I(X^V;Z),\nonumber\\
    \text{subject to}&&\quad -I(X_S;Z)-I(X_{S^c};Z)\leq \eta'_S,\quad\forall S\in\Pi_V,\IEEEeqnarraynumspace\IEEEyesnumber\label{eq:resp_nv_compete_form}
\end{IEEEeqnarray}
where $\eta'_S>0,\forall S\in\Pi_V$ are the control thresholds; $\Pi_V$ denotes the bipartite set of $[V]$ as defined previously. The constraint of \eqref{eq:resp_nv_compete_form} is a straightforward generalization of \eqref{eq:rep_2v_key_relation} that applies to all possible bipartite of the index set $[V]$:
\begin{IEEEeqnarray}{rCl}
    I(X_S;X_{S^c}|Z)=I(X^V;Z)-I(X_S;Z)-I(X_{S^c};Z)+I(X_S;X_{S^c}),\IEEEeqnarraynumspace\IEEEyesnumber\label{eq:resp_nv_key_relation}
\end{IEEEeqnarray}
where the last term $I(X_S;X_{S^c})$ is a constant with respect to $P_\theta(Z|X^V)$ since $P(X^V)=P(X_S,X_{S^c})$ is known \textit{a priori}. The key insight is that the \textit{Bipartite} form focuses on the conditional distributions that satisfy:
\begin{IEEEeqnarray}{rCl}
    P_\theta(X^V|Z) = P_\theta(X_S|Z)P_\theta(X_{S^c}|Z),\quad \forall S\in\Pi_V.\IEEEeqnarraynumspace\IEEEyesnumber\label{eq:bipartite_pxcz_explain}
\end{IEEEeqnarray}
Notably, \eqref{eq:resp_nv_compete_form} relaxes the conditional independence through the defined bipartite set $\Pi_V$ which is in sharp contrast to~\cite[Definition 3]{sula2021common}.
Finally, the complexity of the \textit{Bipartite} common information, defined as the growth rate of the number of parameters with respect to the number of information sources is $\mathcal{O}(|\mathcal{Z}||\mathcal{X}|^V)$ when assuming $|\mathcal{X}_i|=|\mathcal{X}|,\forall i\in[V]$.

\subsection{Variational Common Information}
To address the exponential complexity growth rate of the \textit{Bipartite} form, we further propose the \textit{Variational} form of common information. The high complexity is because of the dimensionality of the variable $P_\theta(Z|X^V)$. According to Bayes' rule, it can be equivalently expressed as:
\begin{IEEEeqnarray}{rCl}
    P_\theta(Z|X^V)=\frac{P_\theta(Z)P_\theta(X^V|Z)}{\sum_{z'\in\mathcal{Z}}P_\theta(z')P_\theta(X^V|z')}=\frac{e^{\log{P_\theta(Z)}+\log{P_\theta(X^V|Z)}}}{\sum_{z'\in\mathcal{Z}}e^{\log{P_{\theta}(z')}+\log{P_\theta(X^V|z')}}}.\IEEEeqnarraynumspace\IEEEyesnumber\label{eq:var_pzcx_softmax}
\end{IEEEeqnarray}
Therefore, one can also let $P_\theta(Z),P_\theta(X^V|Z)$ be the variable to optimize with alternatively and use the above projection to recover solutions of the original feasible set. However, the factorization relation \eqref{eq:bipartite_pxcz_explain} in \textit{Bipartite} form still results in an exponential complexity growth rate. To address this, we further impose the conditional independence, resulting in the following factorization relation:
\begin{IEEEeqnarray}{rCl}
    P_\theta(X^V|Z)=\prod_{i=1}^VP_\theta(X_i|Z).\IEEEeqnarraynumspace\IEEEyesnumber\label{eq:vi_condi_indep}
\end{IEEEeqnarray}
Under this restriction, the complexity growth rate reduces to linear $\mathcal{O}(V|\mathcal{Z}||\mathcal{X}|)$ in the number of sources $V$. However, because of the change of variables, an extra constraint is introduced. To see this, define $P_\theta(X^V)=\sum_{z\in\mathcal{Z}}P_\theta(z)\prod_{i=1}^VP_\theta(X_i|z)$. Here, since both $P_\theta(Z),\{P_\theta(X_i|Z)\}_{i=1}^V$ are variables, $P_\theta(X^V)$ might not equal to $P(X^V)$. In contrast, for the variables considered in \eqref{eq:wyner_2source}, $\sum_{z\in\mathcal{Z}}P_\theta(Z|X^V)P(X^V)=P(X^V)$, i.e., $P_\theta(X^V)$ always matches $P(X^V)$. To maintain the matching conditions, we impose a KL divergence constraint between $P_\theta(X^V)$ and $P(X^V)$. That is, we tolerate a small approximation mismatch between $P_\theta(X^V)$ and $P(X^V)$. This is formulated \textit{Variational Common Information} optimization problem~\cite{huang2023efficient}:
\begin{IEEEeqnarray}{rCl}
    \underset{P_\theta(Z),{\{P_\theta(X_i|Z)\}_{i=1}^V}}{\min}&&\quad I_\theta(X^V;Z),\nonumber\\
    \text{{subject to}}&&\quad D_{KL}[P_\theta(X^V)\parallel P(X^V)]\leq \eta_{X^V},\IEEEeqnarraynumspace\IEEEyesnumber\label{eq:opt_var_form} 
\end{IEEEeqnarray}
where $\eta_{X^V},\eta_{Z}>0$ are some thresholds and $D_{KL}[\mu\parallel \nu]$ denotes the KL divergence between two properly defined measures $\mu,\nu$ in a proper support~\cite{cover1999elements}; and we define $P_\theta(X^V):=\sum_{z\in\mathcal{Z}}P_\theta(Z)\prod_{i=1}^VP_\theta(X_i|Z)$; The subscript $\theta$ indicates that the metrics are computed from the variables $\Theta:=(P_\theta(Z),\{\{P_\theta(X_i|Z)\}_{i=1}^V\})$.

Compared to Wyner's formulation \eqref{eq:wyner_2source}, the difference is that the variables to optimize with are the collection of marginal and conditional probabilities $\Theta$. This parameter space has linear growth rate with respect to the number of sources $V$, since $|\Theta|=\mathcal{O}(V|\mathcal{Z}||\mathcal{X}|)$ when assuming $|\mathcal{X}|=|\mathcal{X}_i|,\forall i\in[V]$, which significantly reduces the complexity growth as compare to the baselines whose growth rate is at $\mathcal{O}(|\mathcal{Z}||\mathcal{X}|^V)$. 
Note that one can map the solution from solving \eqref{eq:opt_var_form} to that of Wyner's formulation through simplex projection \eqref{eq:var_pzcx_softmax}.

\section{Known Joint Distribution}
In this part, we focus on the settings where the joint distribution of the information sources $P(X^V)$ is available, align with Wyner's formulation and most of the recent extensions~\cite{wyner2sources,sula2021gray,kumar2014exact}. Leveraging this knowledge, we develop two efficient solvers for the proposed two extensions of the Wyner common information. For simplicity, we focus on discrete random variables in this section.

\subsection{Solving Bipartite Common Information}\label{subsec:admm_impl}
Starting from the problem statement \eqref{eq:resp_nv_compete_form}. Using Lagrange multipliers to express the optimization problem into an unconstrained form. Then the loss function $\mathcal{L}^{(V)}_\beta$ to be minimized with is given by:

\begin{IEEEeqnarray}{rCl}
    \mathcal{L}^{(V)}_\beta:=I(X^V;Z)+\sum_{S\in\Pi_V}\beta_S\left[I(X_S;X_{S^c}|Z)-\eta_S\right],\IEEEeqnarraynumspace\IEEEyesnumber\label{eq:resp_nv_lag}
\end{IEEEeqnarray}
where $S\in\Pi_V$ is a partition of the index set $[V]$ and $\{\beta_S\}$ is the set of multipliers and $\{\eta_S\}$ is the set of non-negative control thresholds. As discussed in \eqref{eq:resp_nv_key_relation}, $\mathcal{L}_\beta^{(V)}$ is non-convex with respect to $P(Z|X^V)$ (the subscript $\theta$ is omitted since $P(X^V)=P_\theta(X^V)$ always in this case) so we bypass the computation difficulty with relaxations. We focus on a fixed set of multipliers $\{\beta_S\}_{S\in\Pi_V}$, minimize the relaxed objective function and compare the obtained solutions to the control thresholds $\{\eta_S\}_{S\in\Pi_V}$. Then one can search over a grid of $(0,M]^{|\Pi_V|}$ to find solutions that characterize the trade-off.

In minimizing the relaxed objective, the unconstrained Lagrangian \eqref{eq:resp_nv_lag} can be expressed as an equivalent form by substituting \eqref{eq:resp_nv_key_relation} into \eqref{eq:resp_nv_lag}:
\begin{IEEEeqnarray}{rCl}
    \tilde{\mathcal{L}}^{(V)}_\beta=\left(1+\sum_{S\in\Pi_V}\beta_S\right)I(X^V;Z)-\sum_{S\in\Pi_V}\beta_S\left[I(X_S;Z)+I(X_{S^c};Z)\right] +C_\beta,\IEEEeqnarraynumspace\IEEEyesnumber\label{eq:bipartite_loss_clean}
\end{IEEEeqnarray}
where $C_\beta>0$ is a constant independent of $P(Z|X^V)$. Then by re-writing the mutual information as combinations of entropy and conditional entropy functions, we obtain:
\begin{IEEEeqnarray}{rCl}
    \mathcal{L}^{(V)}_\beta=-\left(1+\sum_{S\in\Pi_V}\beta_S\right)H(Z|X^V)+\left(1+\sum_{S\in\Pi_V}\beta_S\right)H(Z)-\sum_{S\in\Pi_V}\beta_S\left[I(Z;X_S)+I(Z;X_{S^c})\right].\IEEEeqnarraynumspace\IEEEyesnumber\label{eq:resp_nv_ent_lag}
\end{IEEEeqnarray}
\subsubsection{Algorithm}
We propose solving \eqref{eq:resp_nv_ent_lag} with Difference of Convex Algorithm (DCA). This is in sharp contrast to previous solvers that do not exploit the difference-of-convex (DC) structure~\cite{sula2021common,huang2023efficient}. It is well-known that DCA can solve DC problems efficiently with guarantees of convergence to local stationary points. A DC optimization problem can be expressed as the following standard form:
\begin{IEEEeqnarray}{rCl}
    &\underset{x\in\mathcal{X}}{\min}\,& f(x)-g(x),\IEEEeqnarraynumspace\IEEEyesnumber\label{eq:dca_basic}
\end{IEEEeqnarray}
where $f,g$ are convex function with respect to $x\in\mathcal{X}$. For simplicity, we assume that $f,g$ are twice differentiable. The idea of DCA is to approximate the concave part $-g(x)$ with the first order expansion and minimizes the obtained surrogate objective. And since $g(x)$ is convex, the surrogate is an upper bound of the original objective function. In other words, DCA ``linearizes'' the concave part of \eqref{eq:dca_basic} then minimizes the convex surrogate upper bounds iteratively:
\begin{IEEEeqnarray}{rCl}
    &x^{k+1}:=\underset{x\in\mathcal{X}}{\arg\min}\,&f(x)-g(x^k)-\langle\nabla g(x^k),x-x^k \rangle,\IEEEeqnarraynumspace\IEEEyesnumber\label{eq:dca_abs_alg}
\end{IEEEeqnarray}
where $k$ denotes the iteration counter and $x^0$ can be randomly chosen from the feasible solution set. For more details about DC programming and DCA, we refer to the excellent review and the references therein that summarize the development of DCA including recent applications~\cite{le2018dc}.

One of the difficulties of adopting DCA is the separation of the objective function. This is due to the infinite number of combinations of convex function pairs to choose from, e.g., $f(x)+\mu\lVert x\rVert^2$ and $g(x)+\mu\lVert x\rVert^2,\forall \mu\in\mathbb{R}^+$, and some of them converge to poor stationary points more often whereas others choices might have slower convergence rate. In the proposed framework, we address the difficulty by using the key relation \eqref{eq:resp_nv_key_relation}, which leads to the following DC separation:
\begin{IEEEeqnarray}{rCl}
    f(P_{z|x^V}):&=&-(1+\sum_{S\in\Pi_V}\beta_S)H(Z|X^V),\IEEEyesnumber\label{eq:dca_fg}\IEEEeqnarraynumspace\IEEEyessubnumber\label{eq:dca_f_nv}\\
    g(P_{z|x^V}):&=&-(1+\sum_{S\in\Pi_V}\beta_S)H(Z)+\sum_{S\in\Pi_V}\beta_S\left[I(Z;X_S)+I(Z;X_{S^c})\right],\IEEEeqnarraynumspace\IEEEyessubnumber\label{eq:dca_g_nv}
\end{IEEEeqnarray}
where $P_{z|x^V}$ denotes the vector form of $p(Z|X^V)$, obtained by cascading the $|\mathcal{Z}|$-dimensional probability vectors given each realization of $x^V\in\mathcal{X}^V$ into a long vector. Both $f,g$ in \eqref{eq:dca_fg} are convex with respect to $P_{z|x^V}$ since the negative (conditional) entropy and mutual information are convex functions with respect to the vector variables for known $P(X^V)$~\cite{huang2022IT,cover1999elements}.

Then by applying DCA \eqref{eq:dca_abs_alg} to \eqref{eq:dca_fg}, followed by an elementary functional derivative analysis (included in Appendix \ref{appendix:wyner_dca_derivation}), we obtain the \textit{Bipartite} solver for \eqref{eq:resp_nv_compete_form}:
\begin{IEEEeqnarray}{rCl}
    &p^{k+1}(z|x^V):=&\frac{p^{k}(z)}{M^k(x^V,\{\kappa_S\}_{S\in\Pi_V})}\exp\left\{\sum_{S\in\Pi_V}\kappa_S\left[\log{\frac{p^{k}(x_S|z)}{p(x_S)}}+\log{\frac{p^k(x_{S^c}|z)}{p(x_{S^c})}}\right]\right\},\IEEEyesnumber\label{eq:wynerdca}\IEEEeqnarraynumspace\IEEEyessubnumber\label{eq:wynerdca_pzcxv}\\
    &p^{k+1}(z):=&\sum_{x^V\in\mathcal{X}^V}p(x^V)p^{k+1}(z|x^V),\IEEEeqnarraynumspace\IEEEyessubnumber\label{eq:wynerdca_pz}\\
    &p^{k+1}(z|x_S):=&\sum_{x_{S^c}\in\mathcal{X}_{S^c}}p(x_{S^c}|x_S)p^{k+1}(z|x^V),\quad\forall S\in\Pi_V,\IEEEeqnarraynumspace\IEEEyessubnumber\label{eq:wynerdca_complement}    
\end{IEEEeqnarray}
where $M^k$ is the step-$k$ normalization function so that $p^{k+1}(z|x^V)$ is a valid conditional probability and $\kappa_S:=\beta_S/(1+\sum_{S'\in\Pi_V}\beta_{S'})$. Note that one can obtain $p^{k}(x_S|z)=p^{k}(z|x_S)p(x_S)/p^k(z),\forall S\in\Pi_V$ by Bayes rule.

\subsubsection{Convergence Analysis}

\begin{theorem}\label{thm:wynerdca_conv}
    The sequence $\{P^k_{z|x^V}\}_{k\in\mathbb{N}}$ obtained from the \textit{Bipartite} solver~\eqref{eq:wynerdca} converges to a local stationary point $w^*$. 
\end{theorem}
\begin{IEEEproof}[proof sketch]
    See Appendix \ref{appendix:pf_thm_wynerdca_conv} for detailed proof. The convergence of the \textit{Bipartite} solver is based on the explicit first-order update: $\nabla f(P_{z|x^V}^{k+1})=\nabla g(P^k_{z|x^V})$, which corresponds to the self-consistent equation~\eqref{eq:wynerdca_pzcxv}. The analysis relies on the descent lemma (see Appendix \ref{appendix:pf_thm_wynerdca_conv}) which assures that the sequence $\{f(P^k_{z|x^V})-g(P^k_{z|x^V})\}_{k\in\mathbb{N}}$, obtained from the \textit{Bipartite} solver has non-increasing objective function values between consecutive updates, hence converging to a stationary point. Combining the fact that the objective function is a non-convex function with respect to $P_{z|x^V}$, the convergence guarantee is qualified to local stationary points.
\end{IEEEproof}
 As a remark, following the common practice in non-convex optimization literature to escape local stationary points~\cite{franti2019much}, one can run the solver multiple times with random initialization and report the best solution with the lowest objective function value as the final result (see Section \ref{subsec:eval_known}).

\subsection{Solving Variational Common Information}
As for solving the \textit{Variational Common Information}~\eqref{eq:opt_var_form}. Recall that $P_\theta(Z,X^V)$ is independent to $P(X^V)$ as mentioned in \eqref{eq:opt_var_form}. Since $P(x^V)$ is known for all realizations $x^V\in\mathcal{X}^V$, it is desirable to include $P(X^V)$ into the computation. Hence, we develop the following surrogate loss upper bound for the mutual information that retains the knowledge of the given $P(X^V)$:
\begin{lemma}[Lemma 1 \cite{huang2023efficient}]\label{lemma:vi_full_know_ub}
Let $P_\theta(X):=\sum_{z\in\mathcal{Z}}p(z)P_\theta(X|z)$ be distinct to $P(X)$ with $Z$ be a measurable random variable, then we have:
\begin{IEEEeqnarray}{rCl}
    I_\theta(Z;X^V)\leq -\sum_{i=1}^VH_\theta(X_i|Z)-\mathbb{E}_{X^V;\theta}\left[\log{P(X^V)}\right],\IEEEeqnarraynumspace\IEEEyesnumber\label{eq:var_full_know_mi_bound}
\end{IEEEeqnarray}
where the equality holds when $P(X^V)=P_\theta(X^V)$. The expectation is taken with respect to $P_\theta(X^V)$.
\end{lemma}
\begin{IEEEproof}
See Appendix \ref{appendix:pf_lemma_vi_full_know_ub} for details. The derivation follows the key relation:
\begin{IEEEeqnarray}{rCl}
    I_\theta(Z;X^V)=-D_{KL}[p_\theta(X^V)\parallel p(X^V)]-\mathbb{E}_{X^V;\theta}\left[\log{P(X^V)}\right]-\sum_{i=1}^VH_\theta(X_i|Z),\IEEEeqnarraynumspace\IEEEyesnumber\label{eq:vi_key_relation}
\end{IEEEeqnarray}
where $H_\theta(X^V|Z)=\sum_{i=1}^VH_\theta(X_i|Z)$ is due to the design of the feasible solution set $\Theta$. 
\end{IEEEproof}

By Lemma \ref{lemma:vi_full_know_ub}, we arrive at the surrogate problem of~\eqref{eq:opt_var_form}:
\begin{IEEEeqnarray}{rCl}
    \underset{P_\theta(Z),\{P_\theta(X_i|Z)\}_{i=1}^V}{\min}\quad&& -\sum_{i=1}^VH_\theta(X_i|Z)-\mathbb{E}_{X^V;\theta}\left[\log{P(X^V)}\right],\nonumber\\
    \text{subject to}\quad&&D_{KL}[P_\theta(X^V)\parallel P(X^V)]\leq \eta'_X,\IEEEeqnarraynumspace\IEEEyesnumber\label{eq:discrete_vi_problem}
\end{IEEEeqnarray}
where $\eta'_X>0$ is a control threshold. Then we can apply a Lagrange multiplier and write \eqref{eq:discrete_vi_problem} into an unconstrained form. It is treated as the loss function $\mathcal{L}_\gamma$ for minimization:
\begin{IEEEeqnarray}{rCl}
    \mathcal{L}_{\gamma}:= -\sum_{i=1}^VH_\theta(X_i|Z)-\mathbb{E}_{X^V;\theta}\left[\log{P(X^V)}\right] +\gamma D_{KL}[P_\theta(X^V)\parallel P(X^V)],\IEEEeqnarraynumspace\IEEEyesnumber\label{eq:am_solver_bound}
\end{IEEEeqnarray}
where $\gamma>0$ is a multiplier. The loss function \eqref{eq:am_solver_bound} is non-convex with respect to the parameter set $\{P_\theta(X_i|Z)\}_{i=1}^V$. This is due to that $P_\theta(X^V)=\sum_{z\in\mathcal{Z}}P_\theta(z)\prod_{i=1}^VP_\theta(X_i|z)$. To avoid the difficulties in non-convex optimization, we adopt the following relaxations. First, we set the multiplier $\gamma$ as a fixed constant during the process of minimization, then we sweep through a range of $\gamma$ to find the minimum loss as the reported solution. Second, we assume $P_\theta(Z)=1/|\mathcal{Z}|$ (this assumption can be relaxed readily without affecting the presented results) instead of jointly optimize $\{P_\theta(X_i|Z)\}_{i=1}^V$, we update an element $P(X_i|Z),\forall i\in[V]$ with others fixed sequentially with respect to the index number. Together, the two steps divide the full problem into a sequence of relaxed convex sub-problems and each of them can be solved efficiently.

\subsubsection{Algorithm}
We propose the following iterative solver to update the parameters corresponding to each information source sequentially. Denote the set of parameterized distributions for the $i^{th}$ source as $l_i:=\{{P(X_i|z)}\}_{z\in\mathcal{Z}}$, the update follows by:
\begin{IEEEeqnarray}{rCl}
    l_i^{k+1}:=\underset{l_i\in\Theta_i}{\arg\min}\,\bar{\mathcal{L}}_{\gamma}(\{l^{k+1}_j\}_{j<i},l_i,\{l^k_m\}_{m>i}),\IEEEeqnarraynumspace\IEEEyesnumber\label{eq:solver_an_iterative}
\end{IEEEeqnarray}
where the superscript $k$ denotes the iteration counter and $\{l_j\}_{j<1}=\{l_m\}_{m>V}=\emptyset$. $\Theta_i$ denotes the set of probability simplice for the $i^{th}$ source, i.e., $\Theta_i:=\{P_\theta(X_i|z)\}_{z\in\mathcal{Z}}$.

From the first-order functional derivative analysis (included in Appendix \ref{appendix:wyner_am_derivation}), we present the \textit{VI} (Variational Inference) solver. The iterative algorithm consists of the following steps. $\forall w\in[W]_i,\forall i\in[V]$:
\begin{IEEEeqnarray}{rCl}
    p^{k+1}_\theta(x_i|z)=\frac{p(x^w)}{M_w(z)}\exp\left\{-D_{KL}[p^k_\theta(x^w|z)\parallel p(x^w|x_i)]+\beta\sum_{x^w\in\mathcal{X}^{W_i}}p^k_\theta(x^w|z)\log{\frac{p(x_i,x^w)}{p^k_\theta(x_i,x^w)}}\right\},\IEEEeqnarraynumspace\IEEEyesnumber\label{eq:wyneram_nv_main}
\end{IEEEeqnarray}
where we denote $[W]_i:=[V]\backslash\{i\}$, $X^{W_i}$ is $X^V$ with $X_i$ excluded. Similarly, we define $\mathcal{X}^{W_i}:=\mathcal{X}^{i-1}_1\times \mathcal{X}_{i+1}^V$. Lastly, we denote $p^k_\theta(x^{W_i}|z):=\prod_{v=1}^{i-1}p^{k+1}_\theta(x_v|z)\prod_{u=i+1}^Vp^k_\theta(x_u|z)$ and $\prod_{i=1}^0=\prod_{i=V+1}^V=1$.

Different from the earlier work~\cite{huang2023efficient}, the \textit{VI} solver has an explicit closed-form update steps~\eqref{eq:wyneram_nv_main} which is simpler to implement. This is in sharp contrast to the gradient-descent-based solvers where the performance heavily depends on a proper step-size through a separate hyperparameter selection phase~\cite{sula2021common,huang2023efficient}.

\subsubsection{Convergence Analysis}
Then we provide the convergence analysis for the proposed solver, which is summarized as follows:
\begin{theorem}
Let $l^k_i$ be defined as in \eqref{eq:solver_an_iterative} where $k$ denotes the iteration counter and $i$ the source index. The sequence $\{l_i^k\}_{k\in\mathbb{N},i\in[V]}$, obtained from the solver \eqref{eq:solver_an_iterative} converges to a local stationary point independent of the initialization.
\end{theorem}
\begin{IEEEproof}[proof]
    See Appendix \ref{appendix:pf_thm_am_solver_conv}. The idea is that each sub-problem is a convex function w.r.t. $\theta_i$ with $\{\theta_j\}_{j\neq i}$ fixed. However, since the full objective function is non-convex w.r.t. $\{\theta_i\}_{i\in[V]}$, the convergence assures a local stationary point only. 
\end{IEEEproof}

\section{Unknown Joint Distribution with Empirical Samples}\label{sec:unknown_dist}
As for more general settings beyond discrete $X^V,Z$, the computation challenge of Wyner common information remains open except for certain jointly Gaussian special cases~\cite{xulossy16,sula2021gray}. Complementing these earlier efforts, we generalize the efficient solvers for \textit{Variational} and \textit{Bipartite} formulations to unknown joint distribution settings but assume availability of a few empirical samples of it.
Without full knowledge of the joint distribution $P(X^V)$, estimating $P_\theta(X^V)$ is prohibitively difficult. Yet, it is relatively easy to extract the relevant common information from the multiple sources of observations directly. The insight of the latter approach is that it is often more efficient with significantly lower computation complexity but still gives good performance. However, estimating common information with empirical samples introduces an additional computational challenge: the efficiency of the estimation process and the fusion of the extracted features from each modal-specific information source. Here, the two questions that we focus on in this section are: Conditioned on the common randomness, how to estimate the statistically independent parts between information sources efficiently with empirical samples? Following this, what are the characteristics of the common randomness when the estimates are given? In answering these questions, we leverage the insights of the proposed \textit{Bipartite} and \textit{Variational} common information and extend them to empirical settings. Moreover, we employ efficient parametric estimation methods for each of the proposed extensions. In each case, we obtain closed-form expressions of the characteristics of the common randomness.
\subsection{Bipartite Common Information Estimation}
Given $N$ empirical samples (a dataset $\mathcal{D}^V_N=\{s^{(i)}|s^{(i)}\sim p(X^V)\}_{i=1}^N$), the \textit{Bipartite} common information in unknown distribution settings can be estimated through the following optimization problem:
\begin{IEEEeqnarray}{rCl}
\underset{\{\theta_S\in\Theta_S\}_{S\in\Pi_V}}{\min}\,&& \frac{1}{N}\sum_{n=1}^N\left\{D_{KL}[P_\theta(Z|x_V^{(n)})\parallel P_\theta(Z)] - \sum_{S\in\Pi_V}\kappa_S P_\theta(Z|x_V^{(n)})\left[\log{\frac{P_\theta(Z|x^{(n)}_S)}{P_\theta(Z)}}+\log{\frac{P_\theta(Z|x^{(n)}_{S^c})}{P_\theta(Z)}}\right]\right\},\nonumber\\
\text{subject to}\,&&P_\theta(Z|x^V)=\frac{\exp\left\{\log{P_\theta(Z)}+\sum_{S\in\Pi_V}\kappa_S\left[\log{\frac{P_\theta(Z|x_S)}{P_\theta(Z)}+\log{\frac{P_\theta(Z|x_{S^c})}{P_\theta(Z)}}}\right]\right\}}{M(x^V,\{\theta_S,\kappa_S\}_{S\in\Pi_V})},\IEEEeqnarraynumspace\IEEEyesnumber\label{eq:unknown_bipartite_prob}
\end{IEEEeqnarray}
where $M$ is a normalization function. It is straightforward to see that \eqref{eq:unknown_bipartite_prob} relates to \eqref{eq:bipartite_loss_clean} when the weak law of large number is applied. The objective function of \eqref{eq:unknown_bipartite_prob} can be estimated with Monte-Carlo sampling, and hence is applicable to problems with a large number of data samples without knowing $P(X^V)$.

\subsubsection{Implementation}\label{subsec:bi_impl}
To solve \eqref{eq:unknown_bipartite_prob}, we first assume a marginal distribution $P_\theta(Z)$ as a reference probability density/mass function, e.g., standard normal $\mathcal{N}(0,I)$ or discrete uniform distribution $\mathcal{U}_Z=1/|\mathcal{Z}|$, then we parameterize the conditional log-likelihoods $\log P_\theta(Z|X_S)$. 
Specifically, conditioned on a $x_S\in\mathcal{X}_S$ the common randomness is distributed according to a probability density/mass function. We adopt the well-known exponential family of distributions for this step as it is a well-known efficient model for characterizing the conditionally independent noise. In the following, we present two examples to demonstrate the effectiveness of the implementation.

Consider the following Gaussian model. Conditioned on a single source sample $\boldsymbol{x}_S\in\mathcal{X}_S$, if $\boldsymbol{z}\sim\mathcal{N}(\boldsymbol{\mu}_{x_S},\boldsymbol{\Sigma}_{x_S})$, then the log-likelihood is given by:
\begin{IEEEeqnarray}{rCl}
    \log{P_\theta(\boldsymbol{z}|\boldsymbol{x}_S)}=-\frac{1}{2}\left[d_z\log{2\pi}+\log|\boldsymbol{\Sigma}_{x_S}|+(\boldsymbol{z}-\boldsymbol{\mu}_{x_S})^T\boldsymbol{\Sigma}_{x_S}^{-1}(\boldsymbol{z}-\boldsymbol{\mu}_{x_S})\right],\IEEEeqnarraynumspace\IEEEyesnumber\label{eq:impl_gaussian_llr}
\end{IEEEeqnarray}
where $d_z$ is the dimension of $\boldsymbol{z}$; $\boldsymbol{\mu}_{x_S}$ is the mean and $\boldsymbol{\Sigma}_{x_S}$ is the covariance matrix, each is parameterized as a function of $\boldsymbol{x}_s$. Here, the conditional noise is parameterized as Gaussian distribution, characterized by its mean and covariance. With the parameters, the log-likelihood can be estimated for each given empirical pair $\{x^{(n)}_S\}_{n=1}^N$ through standard multivariate sampling~\cite{kennedy2021statistical}.

After the estimation of the conditional log-likelihoods for each $x_S$, from the projection for $P_\theta(Z|X^V)$ as defined in \eqref{eq:unknown_bipartite_prob}, the common randomness $Z$ when given the realizations of all information sources is also a Gaussian random variable, i.e.,  $Z|\boldsymbol{x}^V\sim\mathcal{N}(\boldsymbol{\mu}_{eq},\boldsymbol{\Sigma}_{eq})$ where:
\begin{IEEEeqnarray}{rCl}
    \boldsymbol{\mu}_{eq}:=\boldsymbol{\Sigma}_{eq}\left[\sum_{S\in\Pi_V}\kappa_S\left(\boldsymbol{\Sigma}_{x_S}^{-1}\boldsymbol{\mu}_{x_S}+\boldsymbol{\Sigma}_{x_{S^c}}^{-1}\boldsymbol{\mu}_{x_{S^c}}\right)\right],\quad \boldsymbol{\Sigma}_{eq}:=\left[\boldsymbol{I}+\sum_{S\in\Pi_V}\kappa_S\left(\boldsymbol{\Sigma}^{-1}_{x_S}+\boldsymbol{\Sigma}^{-1}_{x_{S^c}}-2\boldsymbol{I}\right)\right]^{-1},\IEEEeqnarraynumspace\IEEEyesnumber\label{eq:impl_gaussian_poe}
\end{IEEEeqnarray}
For the special case where $\kappa_S=1/2|\Pi_V|$, it is also known as the product-of-experts (PoE) in generative multi-modal learning literature~\cite{wu2018multimodalPOE,poe2020feifeiPoE}. Contrary to the references, our derivation is based on the insight of the difference-of-convex structure of the \textit{Bipartite} form, therefore has explicit control $\{\kappa_S\}_{S\in\Pi_V}$ of the correlation of the multi-modal sources. 

The above discussion applies to the exponential family class as distributions. We show this with another example. Consider the mixture of Categorical model with a reference marginal $P_\theta(Z)=1/|\mathcal{Z}|$. The conditional log-likelihood is given by:
\begin{IEEEeqnarray}{rCl}
    \log{P_\theta(z|\boldsymbol{x}_S)}:=\sum_{z\in\mathcal{Z}}\boldsymbol{1}\{z=\hat{z}\}\log{q^{(\hat{z})}_{\boldsymbol{x}_S}},\IEEEeqnarraynumspace\IEEEyesnumber\label{eq:impl_categorical_llr}
\end{IEEEeqnarray}
where $\boldsymbol{1}\{\mathcal{A}\}$ is the indicator function and has value $1$ if the statement $\mathcal{A}$ is true and $0$ otherwise; $\{q^{(\hat{z})}_{x_S}\}_{\hat{z}\in\mathcal{Z}}$ is the parameterized category probabilities and are functions of $X^V$. Before combining the mixture of categorical distributions for characterizing the common randomness, we note that performing a label matching process for each of the categorical distributions is necessary in unsupervised settings. Here, we assume that a standard labeling matching is done~\cite{labelmatching}, then by the projection equation \eqref{eq:unknown_bipartite_prob}, the common randomness is also a categorical distribution with the equivalent category probability functions given by:
\begin{IEEEeqnarray}{rCl}
    q^{(z)}_{eq}:=\frac{\prod_{S\in\Pi_V}\left[q_{x_S}^{(z)}q_{x_{S^c}}^{(z)}\right]^{\kappa_S}}{\sum_{z'\in\mathcal{Z}}\prod_{\pi\in\Pi_V}\left[q_{x_{\pi}}^{(z')}q_{x_{\pi^c}}^{(z')}\right]^{\kappa_\pi}}=\textit{Softmax}\left(\sum_{S\in\Pi_V}\kappa_S\left[\log{q^{(z)}_{x_S}}+\log{q^{(z)}_{x_{S^c}}}\right]\right),\quad \forall z\in\mathcal{Z}.\IEEEeqnarraynumspace\IEEEyesnumber\label{eq:impl_categorical_qprob}
\end{IEEEeqnarray}
As for more general forms that include members of the exponential family, such as Bernoulli, Poisson and exponential distributions, we provide a generic model in Appendix \ref{appendix:exp_comm_info}. Finally, the KL divergence in the objective functions \eqref{eq:unknown_bipartite_prob} has closed-form expression when the two distributions belong to the same member class of the exponential family. For instance, in the Gaussian mixture model, since the reference prior distribution $P_\theta(Z)$ and the common randomness (conditioned on $x^V$) are Gaussian distributions, the KL divergence is well-known to be:
\begin{IEEEeqnarray}{rCl}
    D_{KL}[\mathcal{N}(\boldsymbol{\mu}_{eq},\boldsymbol{\Sigma}_{eq}\parallel \mathcal{N}(\boldsymbol{0},\boldsymbol{I}))]=-\frac{1}{2}\left[\log{|{\boldsymbol{\Sigma_{eq}}}|}-d_z+Tr\{\boldsymbol{\Sigma}^{-1}_{eq}\}+\boldsymbol{\mu}_{eq}^T\boldsymbol{\Sigma}_{eq}^{-1}\boldsymbol{\mu}_{eq}\right].\IEEEeqnarraynumspace\IEEEyesnumber\label{eq:impl_gaussian_dkl}
\end{IEEEeqnarray}
This completes the implementation of the \textit{Bipartite} form. In optimizing the parameters, one can adopt amortized representation learning, e.g., parameterized Gaussian means and variances as functions of the observations~\cite{kingma2013auto}, compute the objective function value from a mini-batch of empirical samples, and optimize the parameters with gradient-descent. 

Remarkably, the proposed \textit{Bipartite} form \eqref{eq:unknown_bipartite_prob} is in sharp contrast to the generalized evident lower bound in recent multi-modal learning literature~\cite{sutter2021generalized,xu2021multivae,poe2020feifeiPoE,hwang2021multi}. Here, we focus on the characterization of the common randomness between multi-modal observations, bypassing the computation burden of estimating the unknown $P(X^V)$. Our projection step $P_\theta(Z|X^V)$ follows the insight of the difference-of-convex structure of the formulated \textit{Bipartite Common Information}, allowing the direct estimation of the mutual information, whereas the references rely on the theory of variational inference to obtain a surrogate upper bound of the mutual information, which is essentially a maximum likelihood estimate of the unknown distribution $P_\theta(X^V)\approx P(X^V)$. 

\subsubsection{Correlation of Extracted Features}
 Previously, we made no assumption on the information sources except the conditional independence. In practice, to avoid the ``curse of dimensionality''~\cite{indyk1998approximate}, the observations from the information sources $\{x_{[V]}^{(n)}\}_{n=1}^N$ are pre-processed with dimension reduction techniques such as principal component analysis, spectral decomposition, or autoencoders~\cite{MvCsurvey2018}. Following this perspective, we consider a pre-processed dataset $\{w_{[V]}^{(n)}\}_{n=1}^N$ instead of the original one. However, in contrast to most of the earlier works in multi-modal learning, this data-preprocessing phase is part of the proposed frameworks as well. To see this, recall the key relation in \textit{Bipartite} form:
 \begin{IEEEeqnarray}{rCl}
     I(W_S;W_{S^c}|Z)=I(Z;W^V)-I(Z;W_S)-I(Z;W_{S^c})+I(W_S;W_{S^c}).\IEEEeqnarraynumspace\IEEEyesnumber\label{eq:data_preprocessing_key_relation}
 \end{IEEEeqnarray}
 When the original sources $\{x_{[V]}^{(n)}\}_{n=1}^N$, $I(X_S;X_{S^c})$ is a constant independent to the parameters $\theta\in\Theta$. However, when the pre-processed dataset $\{w_{[V]}^{(n)}\}_{n=1}^N$ is considered instead, the mutual information $I_\phi(W_S;W_{S^c})$ is an objective function to optimize with by itself. This will affect the correlations between extracted source-features. Intuitively, one would expect maximizing $I_\phi(W_S;W_{S^c})$, yet it is important to note that $I_\phi(W_S;W_{S_c})$ is a component of the conditional mutual information. Therefore, minimizing these terms keeps features that satisfy conditional independence. In Section \ref{sec:applications}, we apply this to multi-modal clustering as the correlation optimization sub-objective of the proposed frameworks. Combining the intuition with the conditional mutual information constraints in the \textit{Bipartite} common information, we obtain a unified objective function:
 \begin{IEEEeqnarray}{rCl}
     \underset{\theta\in\Theta}{\min}\,\underset{\phi\in\Phi}{\min}\, I_\theta(Z;X^V)+\sum_{S\in\Pi_V}\kappa_S\left[-I_\theta(Z;X_S)-I_\theta(Z;X_{S^c})+I_\phi(W_S;W_{S^c})\right].\IEEEeqnarraynumspace\IEEEyesnumber\label{eq:max_mi_preprocess}
 \end{IEEEeqnarray}
 Observe that the inner problem only affects $\sum_{S\in\Pi_V}\kappa_SI_\phi(X_S;X_{S^c})$, so we can apply mutual information estimation techniques to keep most of the correlation between the pair of sources~\cite{tian2020contrastive,viola1997alignment,maes1997multimodality,hjelm2018learning,poole2019variational}. We provide Theorem \ref{thm:multimodal_mi_up_bound} as a concrete example in the sequel.
 
\subsection{Variational Common Information Estimation}\label{subsec:vi_unknown}
As for the \textit{Variational} form for unknown distribution settings, in addition to the bipartite set $\Pi_V$ condition, we further impose conditional independence $P(X_S|Z)=\prod_{i\in S}P(X_i|Z)$. Therefore, we have:
\begin{IEEEeqnarray}{rCl}
    \sum_{S\in\Pi_V}\kappa_S\left[\log{P_\theta(X_S|Z)}+\log{P_\theta(X_{S^c}|Z)}\right]=\left(\sum_{S\in\Pi_V}\kappa_S\right)\sum_{i=1}^V \log{P_\theta(X_i|Z)},\IEEEeqnarraynumspace\IEEEyesnumber\label{eq:unknown_vi_relation_bi}
\end{IEEEeqnarray}
where $\kappa_S$ is defined in \eqref{eq:wynerdca}. In words, the Variational form is a special case of \eqref{eq:unknown_bipartite_prob}. But under the conditional independence, the problem is reduced to:
\begin{IEEEeqnarray}{rCl}
    \underset{\theta_V\in\Theta_V}{\min}\,&&\frac{1}{N}\sum_{n=1}^N\left\{D_{KL}[P_\theta(Z|x_V^{(n)})\parallel P_\theta(Z)] - \kappa_{[V]}\sum_{i=1}^V P_\theta(Z|x_V^{(n)})\log{\frac{P_\theta(Z|x^V)}{P_\theta(Z)}}\right\},\nonumber\\
\text{subject to}\,&&P_\theta(Z|X^V)=\frac{\exp\left\{\log{P_\theta(Z)}+\kappa_{[V]}\sum_{i=1}^V\log{\frac{P_\theta(Z|X_i)}{P_\theta(Z)}}\right\}}{M(X^V,\theta_V)},\IEEEeqnarraynumspace\IEEEyesnumber\label{eq:unknown_vi_prob}
\end{IEEEeqnarray}
for some $\kappa_{[V]}\in(0,1)$. Remarkably, when conditional independence is assumed, the growth rate of the complexity, defined as the number of parameters required, is linear in the number of sources $\mathcal{O}(V)$. This aligns with the insight from known distribution settings. Yet, as compared to the \textit{Bipartite} form, the capability to capture the correlation between bipartite sources is worse as now the parametric distributions $P_\theta(Z|X_i),\forall i\in[V]$ depends on individual source only. The implementation of the \textit{Variational} form \eqref{eq:unknown_vi_prob} is similar to that presented in Section \ref{subsec:bi_impl}, with the summation of bipartite replaced with summation of the $V$ sources. Similarly, the projection and the coefficient $\kappa_{[V]}$ follow that defined in \eqref{eq:unknown_vi_prob}, and hence the details are referred to Section \ref{subsec:bi_impl}.

The reduced complexity of the Variational form benefits the optimization of correlations between pairwise features as well. Consider the key relation~\eqref{eq:resp_nv_key_relation} where the additional conditions $X_i\rightarrow Z\rightarrow X_j,\forall i\neq j\in[V]$ holds. In this case we have:
\begin{IEEEeqnarray}{rCl}
    I(X_S;X_{S^c}|Z)&=&I(Z;X^V)-I(Z;X_S)-I(Z;X_{S^c})+I(X_S;X_{S^c})\nonumber\\
    &=&I(Z;X^V)-\sum_{i=1}^VI(Z;X_i)+I(X_S;X_{S^c}),\IEEEeqnarraynumspace\IEEEyesnumber\label{eq:vi_reduction_unknow_mimerge}
\end{IEEEeqnarray}
where conditional independence implies that $H(X_S|Z)=\sum_{i\in S}H(X_i|Z)$. We again recover the linear complexity for the \textit{Variational} form as shown in known distribution settings. As for the terms for data-preprocessing, if we further assume that $\kappa_S=\kappa_{[V]},\forall S\in\Pi_V$, then we have:
\begin{IEEEeqnarray}{rCl}
    \sum_{S\in\Pi_V}\kappa_{S}I(X_S;X_{S^c})=\kappa_{[V]}\sum_{i=1}^V\sum_{j=2}^{i-1}\sum_{L\in\mathcal{P}([V]\backslash\{i,j\})}I(X_i;X_j|X_L),\IEEEeqnarraynumspace\IEEEyesnumber\label{eq:vi_reduction_unknow}
\end{IEEEeqnarray}
where $\mathcal{P}(S)$ denotes the power set of $S$. The above expression can be further reduced under the approximation $I(X_i;X_j|X^{W})\approx I(X_i;X_j),\forall W\subseteq[V]\backslash\{i,j\}$. In this case, there are only ${V\choose 2}$ of terms after rearrangement, resulting in a complexity growth rate at $\mathcal{O}(|\mathcal{X}|^2V^2)$.

\section{Application to Multi-Modal Clustering}\label{sec:applications}

For applications, we focus on multi-modal clustering problems. Here we consider the case where each sample of a dataset consists of multiple modalities of observations that come in pairs. Therefore, it follows that the multi-modal observations of each sample share a common randomness, corresponding to the cluster target. For instance, the same digit can be spoken and written but they all represent the same digit. The assumption that the multi-modal observations come in pairs is interpreted as conditioned on the target variable, the multi-modal observations are statistically independent. Following this perspective, it is natural to make the cluster variable $Z$ be the common randomness in Wyner common information. Note that the cardinality of $Z$ is also a variable to optimize with, which can be achieved through applying standard number of cluster detection algorithms in vector quantization or clustering literature for identification. The basic steps for this are: fix a cardinality and run a clustering algorithm until the objective function value converges, repeating the process over a range of cardinalities and record the losses. Finally, the minimum cardinality with a loss lower than a predetermined threshold is the estimated optimal number of clusters. In the following, we instead assume knowing the optimal number of clustering, but refer to~\cite{1994Cdetect1,2010Cdetect2,syakur2018integration} for well-known and recent results in cluster number identification.

As mentioned in Section \ref{sec:unknown_dist}, in multi-modal clustering, it is computationally more efficient to leverage the common randomness than estimating the unknown joint distribution. Both the proposed \textit{Bipartite} and \textit{Variational} forms fit the task well. Given an unlabeled multi-modal training dataset $D_N=\{x_i^V\}_{i=1}^N$ of $N$ samples, we separate the task into three steps:
\begin{enumerate}
    \item We reduce the dimension of each data modality with modality-specific autoencoders, e.g., $W_i={enc}_\theta(X_i),\hat{X}_i={dec}_\theta(W_i)$ $|\mathcal{W}_i|<|\mathcal{X}_i|,\forall i\in[V]$ with parameters $\theta\in\Theta$. This corresponds to estimating $I_\theta(Z;X_i)=H(X_i)-H_\theta(X_i|Z)$ ($i=S,\forall S\in\Pi_V$ for the Bipartite and $i\in[V]$ for the Variational solver) with standard reconstruction objective functions to train the autoencoders, e.g., Mean-Squared-Error (MSE) or Binary Cross-Entropy (BCE).
    \item In addition to reconstruction sub-objectives, we estimate the mutual information $I_\phi(W_i;W_j)$ ($i,j\in[V]$ for the \textit{Variational} form and $i,j\in\Pi_V$ for the \textit{Bipartite} form) with a separate set of parameters $\phi\in\Phi$.
    \item We parameterize the log-likelihoods $\log P_\psi(W_i|Z)$ with the categorical model as presented in \eqref{eq:impl_categorical_llr} with projection given by \eqref{eq:impl_categorical_qprob}.
\end{enumerate}
  Each step is relevant to the multi-modal clustering task: dimension reduction avoids degradation of clustering performance and improves efficiency as in standard clustering literature~\cite{becht2019dimensionality}; We leverage the assumption that the multi-modal observations come in pairs by optimizing the pairwise correlations of the extracted features, enhancing computation efficiency; The insights from the proposed frameworks, implemented in the categorical model~\eqref{eq:impl_categorical_qprob} provide a theoretic-sound approach for the alignment of the extracted modal-specific features to characterize the common randomness, i.e., the final cluster prediction. 

To further demonstrate the strength of the proposed solvers, we provide a concrete application for the correlation optimization of pairwise multi-modal data. In estimating $I_\phi(W_1;W_2)$ (applies to arbitrary pairs or bipartite without loss of generality), we have the following result:
\begin{theorem}\label{thm:multimodal_mi_up_bound}
    For $N$ pairs of samples $S_N=\{(w^{(n)}_1,w^{(n)}_2)\}_{n=1}^N$, whose empirical joint distribution is given by $P_N(W_1,W_2)=\boldsymbol{1}\{i=j|(w_1^{(i)},w_2^{(j)},i,j\in[N])\}$ and $P_N(W_1)=P_N(W_2)=1/N$, the mutual information has the following upper bound:
    \begin{IEEEeqnarray}{rCl}
        I_N(W_1;W_2)\leq -\mathbb{E}_{W^{N}_1}\left[\log{Q_N(W_2^{N}|W_1^{N})}\right]:=\mathcal{L}^{1,2}_{cor},\IEEEeqnarraynumspace\IEEEyesnumber\label{eq:corr_loss_def}
    \end{IEEEeqnarray}
    where $Q_N$ is a parameterized conditional distribution corresponds to the pairs of $(w^{(i)}_1,w^{(j)}_2),\forall i=j\in[N]$. And the $Q_N^*$ that achieves the tightest bound is given by:
    \begin{IEEEeqnarray}{rCl}
        Q_N^*:=\underset{\{h_+,h_-\}\in\mathcal{H}}{\arg\sup}\,\frac{1}{N}\sum_{n=1}^N\log{\frac{h_{+}(w_1^{(n)},w_2^{(n)})}{h_{+}(w_1^{(n)},w_2^{(n)})+\sum_{k\neq k}^{N}h_{-}(w_1^{(n)},w_2^{(k)})}},\IEEEeqnarraynumspace\IEEEyesnumber\label{eq:corr_q_star}
    \end{IEEEeqnarray}
    where $h_+,h_-$ are some functional with non-negative ranges over a parametric space $\mathcal{H}$ with $h_+$ corresponds to the sample that comes in pairs whereas $h_-$ are for others.
\end{theorem}
\begin{IEEEproof}
    See Appendix \ref{sec:appendix_mi_min}.
\end{IEEEproof}
In practice, one can parameterize $h_+,h_-$ as neural networks to compute a non-negative correlation metric (correlation optimization modules). For instance, if $h$ computes the cosine-correlation between two neural network outputs $g_\theta(w_1),g_\theta(w_2)$ for $N$ batch of samples, then \eqref{eq:corr_q_star} is equivalent to parameterize $Q_N$ as a categorical distribution with $N$ clusters with $\theta\in\Theta$. Following this, \eqref{eq:corr_loss_def} can be interpreted as the KL divergence between a one-hot categorical distribution (only a single element equals to $1$, corresponding to the index of the sample that come in pairs, and $0$ otherwise) and $Q_N$. In our settings, this ``positive'' sample is easy to identify, which is the pair of extracted multi-modal features. Accordingly, $\phi\in\Phi$ can be optimized by minimizing \eqref{eq:corr_loss_def}. The insight applies to both the proposed methods. For \textit{Bipartite} form, the inputs of the correlation optimization modules refer to the bipartite pairs $(w^{(n)}_S,w^{(n)}_{S^c})$ whereas they are any pair of the modality observations, i.e., $(w^{(n)}_i,w^{(n)}_j), i\neq j\in[V]$, in the \textit{Variational} form. Finally, two remarks are in order:
\begin{itemize}
    \item Theorem \ref{thm:multimodal_mi_up_bound} is similar to the contrastive predictive coding heuristic that maximizes a mutual information lower bound~\cite{oord2018representation,poole2019variational,tian2020contrastive}. In contrast, we obtain an upper bound by leveraging the fact that the multi-modal observations come in pairs, following the proposed common information frameworks.
    \item While the contrastive approach has been adopted in multi-modal learning literature~\cite{xu2022MFLVC,tian2020contrastive} as a heuristic sub-objective, it is part of the proposed common information frameworks.
\end{itemize}

\begin{algorithm}[!b]
\caption{Bipartite Wyner Multi-Modal Clustering}
\label{alg:pseudo_bipartite}
\begin{algorithmic}[1]
\REQUIRE Number of modalities $V$, Dataset $\mathcal{D}^V_N=\{(x^V)^{(i)}\}_{i=1}^N$, Number of Clusters $|\mathcal{Z}|$, Iteration limit $M$.
\ENSURE Cluster distribution $\boldsymbol{q}^*=(q_1,\cdots,q_{|\mathcal{Z}|})$.  
\STATE Initialize autoencoders $\{enc^{(i)}_\theta,dec^{(i)}_\theta\}_{i=1}^V$,\COMMENT{Initialization}
\STATE Initialize Correlation optimization modules $\{\phi_S\}_{S\in\Pi_V}$
\STATE Initialize Categorical parameters $\{\boldsymbol{q}^S_{|\mathcal{Z}|}\}_{S\in \Pi_V}$, set counter $k\leftarrow 0$.
\WHILE{$k<M$}

\FORALL{A mini-batch of $B$ samples $x^V_B\in\mathcal{D}_N^V$}
\STATE $\{w^V_B\} \gets \{enc^{(i)}_\theta(\{x^V_B\})\}_{i=1}^V$, and $\{\hat{x}^{(i)}_B\}_{i=1}^V\gets \{dec_\theta^{(i)}(\{w^V_B\})\}_{i=1}^V$,\COMMENT{Dimension Reduction}
\STATE Compute $\{\mathcal{L}_{MSE}^{(i)}(\{x_{i,B}\},\{\hat{x}_{i,B}\})\}_{i=1}^V$, \COMMENT{Reconstruction Losses}
\IF{In Training Mode}
\STATE $\{h^S_B\}_{S\in\Pi_V}\gets \{\phi_S(\{w^S_B\},\{w^{S^c}_B\})\}_{S\in\Pi_V}$,
\STATE Compute Correlation maximizers $\{\mathcal{L}_{cor}^{S}
(h^S_B,h^{S^c}_B)\}_{S\in\Pi_V}$,\COMMENT{Equation \eqref{eq:corr_q_star}}
\ENDIF
\STATE Compute modal-specific cluster distributions $\{\boldsymbol{q}^S_{|\mathcal{Z}|}(\{w^S_B\})\}_{S\in\Pi_V}$,\COMMENT{Equation \eqref{eq:impl_categorical_llr}}
\STATE Compute final prediction $\boldsymbol{q}^*(\{\boldsymbol{q}^S_{|\mathcal{Z}|}\}_{S\in\Pi_V})$,\COMMENT{Equation \eqref{eq:impl_categorical_qprob}}
\STATE Compute $\mathcal{L}_{MI}= \frac{1}{B}\sum_{B}D_{KL}[\boldsymbol{q}^*\parallel \frac{1}{|\mathcal{Z}|}]$.
\IF{In Training Mode}
\STATE Update weights $\textit{back-propagate}(\{\mathcal{L}^{(i)}_{MSE}\}_{i=1}^V,\{\mathcal{L}^{S}_{cor}\}_{S\in\Pi_V},\mathcal{L}_{MI},\{\kappa_S\}_{S\in\Pi_V})$,\COMMENT{$\kappa_S$ defined in \eqref{eq:wynerdca}}
\ENDIF
\ENDFOR
\STATE $k\gets k+1$ (Training mode) or $k\gets M$ (Inference mode).
\ENDWHILE
\end{algorithmic}
\end{algorithm}

\begin{algorithm}[!b]
\caption{Variational Wyner Multi-Modal Clustering}
\label{alg:pseudo_wvae}
\begin{algorithmic}[1]
\REQUIRE Number of modalities $V$, Dataset $\mathcal{D}^V_N=\{(x^V)^{(i)}\}_{i=1}^N$, Number of Clusters $|\mathcal{Z}|$, Iteration limit $M$.
\ENSURE Cluster distribution $\boldsymbol{q}^*=(q_1,\cdots,q_{|\mathcal{Z}|})$.  
\STATE Initialize autoencoders $\{enc^{(i)}_\theta,dec^{(i)}_\theta\}_{i=1}^V$,\COMMENT{Initialization}
\STATE Initialize Correlation optimization modules $\{\phi_i\}_{i=1}^V$
\STATE Initialize Categorical parameters $\{\boldsymbol{q}^{(i)}_{|\mathcal{Z}|}\}_{i=1}^V$, set counter $k\leftarrow 0$.
\WHILE{$k<M$}

\FORALL{A mini-batch of $B$ samples $x^V_B\in\mathcal{D}_N^V$}
\STATE $\{w^V_B\} \gets \{enc^{(i)}_\theta(\{x^V_B\})\}_{i=1}^V$, and $\{\hat{x}^{(i)}_B\}_{i=1}^V\gets \{dec_\theta^{(i)}(\{w^V_B\})\}_{i=1}^V$,\COMMENT{Dimension Reduction}
\STATE Compute $\{\mathcal{L}_{MSE}^{(i)}(\{x_{i,B}\},\{\hat{x}_{i,B}\})\}_{i=1}^V$, \COMMENT{Reconstruction Losses}
\IF{In Training Mode}
\STATE $\{h^{(i)}_B\}_{i=1}^{V}\gets \{\phi_i(\{w^{(i)}_B\})\}_{i=1}^V$,
\STATE Compute Correlation maximizers $\{\mathcal{L}_{cor}^{(i)}
(h_B^{(i)},h^{(j))}_B)\}_{i\neq j\in[V]}$,\COMMENT{Equation \eqref{eq:vi_reduction_unknow},\eqref{eq:corr_q_star}}
\ENDIF
\STATE Compute modal-specific cluster distributions $\{\boldsymbol{q}^{(i)}_{|\mathcal{Z}|}(\{w^{(i)}_B\})\}_{i=1}^V$,\COMMENT{Equation \eqref{eq:impl_categorical_llr},\eqref{eq:unknown_vi_relation_bi}}
\STATE Compute final prediction $\boldsymbol{q}^*(\{\boldsymbol{q}^{(i)}_{|\mathcal{Z}|}\}_{i=1}^V)$,\COMMENT{Equation \eqref{eq:unknown_vi_prob}}
\STATE Compute $\mathcal{L}_{MI}= \frac{1}{B}\sum_{B}D_{KL}[\boldsymbol{q}^*\parallel \frac{1}{|\mathcal{Z}|}]$.
\IF{In Training Mode}
\STATE Update weights $\textit{back-propagate}(\{\mathcal{L}^{(i)}_{MSE}\}_{i=1}^V,\{\mathcal{L}^{(i,j)}_{cor}\}_{i\neq j\in[V]},\mathcal{L}_{MI},\kappa_{[V]})$,
\ENDIF
\ENDFOR
\STATE $k\gets k+1$ (Training mode) or $k\gets M$ (Inference mode).
\ENDWHILE
\end{algorithmic}
\end{algorithm}
\begin{figure*}[!t]
    \centerline{
        \subfloat[Bipartite Solver]{
            \includegraphics[width=3.2in]{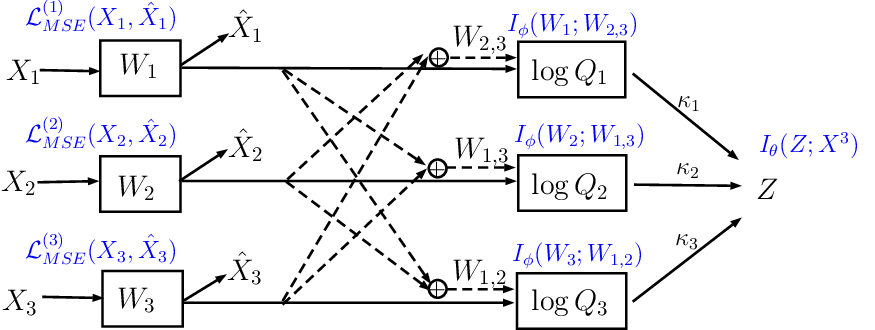}
            \label{subfig:bi_diagram}
        }
        \hfil
        \subfloat[Variational Solver]{
            \includegraphics[width=3.2in]{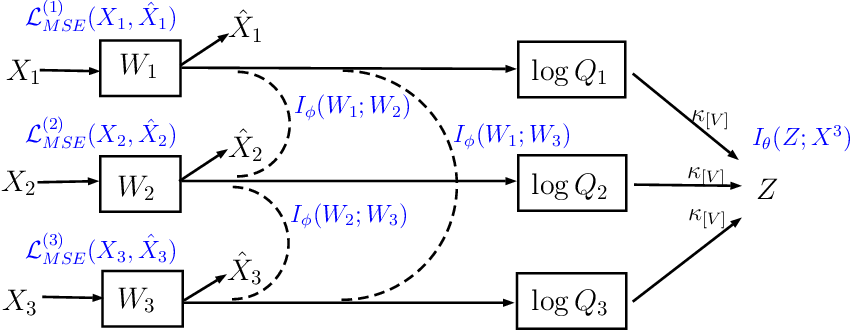}
            \label{subfig:vi_diagram}
        }}
        \caption{Block diagrams of the proposed \textit{Bipartite} and \textit{Variational} solvers in unknown distribution settings with $V=3$. The solid lines represent the feed-forward flow whereas the dotted lines correspond to flows for correlation optimization modules. The computed sub-objective functions are written in blue.}
        \label{fig:block_diagram}
\end{figure*}
For completeness, the pseudocodes of the proposed \textit{Bipartite} and \textit{Variational} solvers in unknown joint distribution settings are shown in Algorithm \ref{alg:pseudo_bipartite} and \ref{alg:pseudo_wvae} respectively. Additionally, the flow charts of the two solvers for the $V=3$ special cases are shown in Fig. \ref{fig:block_diagram} to better explain the three main steps. In Section \ref{subsubsec:nn_impl} we implement the algorithm with neural networks and evaluate them with open-source datasets.

\section{Numerical Results}

\subsection{Evaluation for Known Joint Distribution}\label{subsec:eval_known}
\subsubsection{Evaluated Distributions}
The joint distribution we considered satisfies the conditional independence condition. Specifically, denote the ground-truth variable as $Y$ with the marginal probability $P(y)=1/|\mathcal{Y}|,\forall y\in\mathcal{Y}$, the conditional probability $p(X_i|Y),\forall i\in[V]$ is given by:
\begin{IEEEeqnarray}{rCl}
    p(X_i|Y)=\begin{bmatrix}
        \boldsymbol{l}-\boldsymbol{\delta} & \boldsymbol{0} & \cdots &\boldsymbol{0} & \boldsymbol{\delta}\\
        \boldsymbol{\delta} & \boldsymbol{l}-\boldsymbol{\delta}&\boldsymbol{0} &\cdots & \boldsymbol{0}\\
        \boldsymbol{0} & \boldsymbol{\delta} &\ddots &\ddots & \vdots\\
        \vdots & \ddots & \ddots & \boldsymbol{l}-\boldsymbol{\delta} &  \boldsymbol{0}\\
        \boldsymbol{0} & \cdots & \boldsymbol{0} & \boldsymbol{\delta} & \boldsymbol{l}-\boldsymbol{\delta}
    \end{bmatrix}
\end{IEEEeqnarray}
where $\boldsymbol{l},\boldsymbol{\delta}$ are vectors that will be defined next. Then the joint distribution can be computed as $P(X^V):=\sum_{y\in\mathcal{Y}}p(y)\prod_{i=1}^Vp(X_i|y)$ then given to all the evaluated common information solvers (without access to $Y$). In details, we let $|\mathcal{Y}|=8$, $\boldsymbol{l}=\begin{bmatrix} 0.5 & 0.5\end{bmatrix}^T$; and we consider two cases: \romannum{1}) $\boldsymbol{\delta}_1=\begin{bmatrix} 0 & 0\end{bmatrix}^T$ and \romannum{2}) $\boldsymbol{\delta}_2=\begin{bmatrix} 0.05 & 0.05\end{bmatrix}^T$, resulting in $|\mathcal{X}_i|=16,\forall i\in[V]$. For case \romannum{1}), the realizations of the observations $x_i\in\mathcal{X}_i$ correspond to a single, non-overlap $y\in\mathcal{Y}$ and we call it an ``invertible'' case; For case \romannum{2}) there is ambiguity since a $x_i\in\mathcal{X}_i$ might come from one of the two possible $y\in Y$ (non-invertible). Finally, for a more complete evaluation, the number of observation sources we considered is $V\in\{2,3\}$.  

\subsubsection{Compared Methods}\label{subsubsec:cmp_methods}
We compare the following solvers in this part:
\begin{enumerate}[i)]
    \item Proposed Bipartite~\eqref{eq:wynerdca}: The iterative solver follows the set of self-consistent equations~\eqref{eq:wynerdca} with the parameters $P_\theta(Z|X^V)$. We simplify the hyperparameter set $\{\kappa_S\}_{S\in\Pi_V}$ to a single one $\kappa=\kappa_S,\forall S\in\Pi_V$ and hence $\kappa=1/|\Pi_V|$ as defined in \eqref{eq:wynerdca}.
    \item Proposed VI~\eqref{eq:wyneram_nv_main}: The iterative solver follows the update equation~\eqref{eq:wyneram_nv_main}. The parameters of this solver are the set of conditional probabilities $\{P_\theta(X_i|Z)\}_{i=1}^V$ if $P_\theta(Z)=1/|\mathcal{Z}|$. An iteration of the solver consists of $V$ updates, followed by a projection step~\eqref{eq:var_pzcx_softmax} to obtain a solution $P_\theta(Z|X^V)$ in Wyner's formulation.
    
    \item Baseline~\cite{sula2021common}: The state-of-the-art solver for the \textit{Relaxed} common information~\cite{sula2021gray}. For $V>2$ sources, it minimizes the problem formulated in~\cite[Sec. 7]{sula2021common}. We follow the algorithm, including the choice of hyperparameters, e.g., gradient step size, loss weighting coefficients, etc., adopted in the reference~\cite[Algorithm 1]{sula2021common}. The parameters of this solver are the conditional probability $P_\theta(Z|X^V)$.
\end{enumerate}

Each solver is initialized as follows: we randomly sample from a uniformly distributed source with the range $(0,1)$ then we normalize it to obtain a valid probability vector (or matrix) as the starting point. 

All solvers are evaluated in the same settings. A geometrically spaced grid ranging from $[0.1,10]$ of $20$ points is given as the hyperparameter set. For each element of the set, we initialize a solver as mentioned in the last paragraph and run the solver until a termination condition is reached. This procedure is repeated for $25$ times. The termination condition is either the loss decrease of the solver is less than a threshold $10^{-6}$ or a maximum number of iterations $10^{4}$ is reached.

The output of each solver is a stochastic mapping $P_\theta(Z|X^V)$, which is recorded along with the obtained loss value when terminated. Following this, since $P(X^V)$ is known here, the joint distribution $P_
\theta(Z,X^V)$ is available, so the metrics such as mutual information and (conditional) entropy can be computed readily.

\subsubsection{Solutions to Wyner's Formulation}
\begin{figure*}
    \centerline{
        \subfloat[$V=2,|\mathcal{Y}|=|\mathcal{Z}|=8$ invertible case]{
            \includegraphics[width=3.0in]{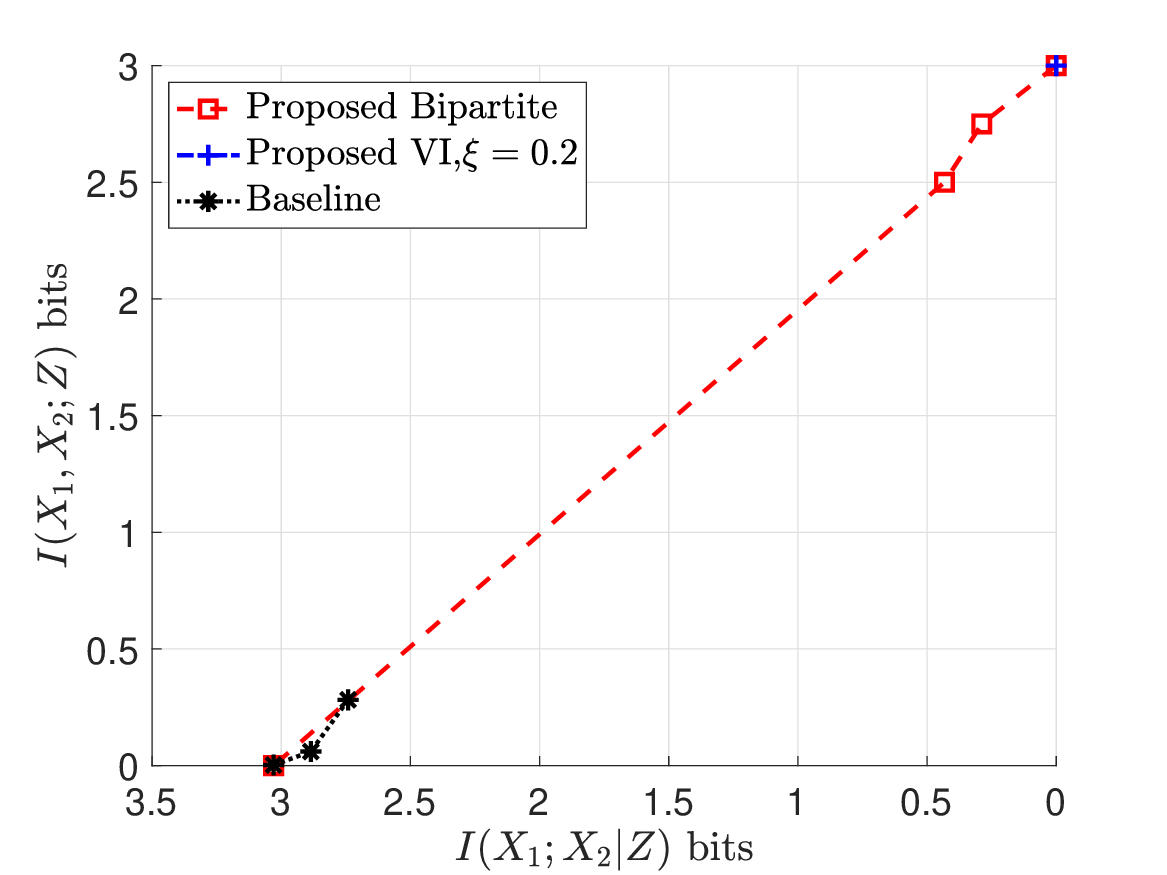}
            \label{subfig:micurve_v2y8c00}
        }
        \hfil
        \subfloat[$V=2,|\mathcal{Y}|=|\mathcal{Z}|=8$ non-invertible case]{
            \includegraphics[width=3.0in]{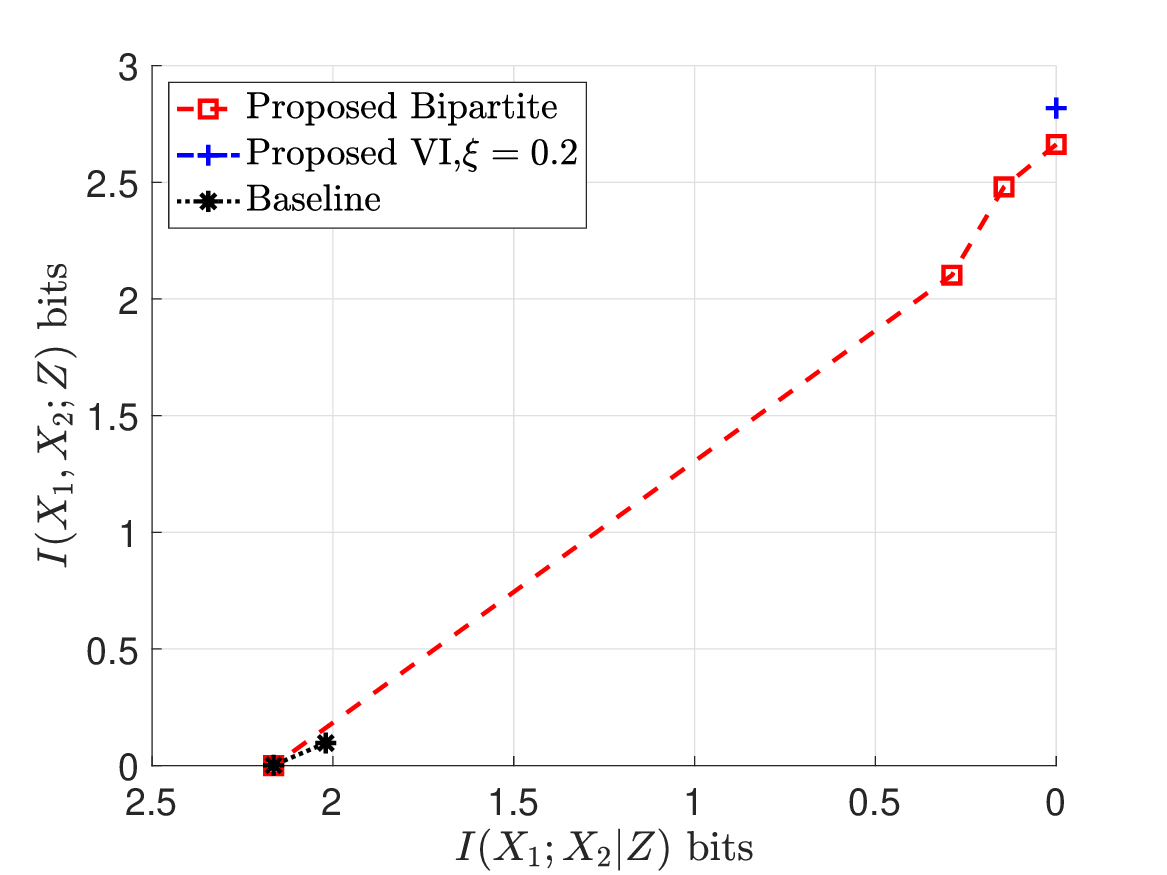}
            \label{subfig:micurve_v2y8c01}
        }}
        \caption{Evaluation results on the information plane. The minimum mutual information $I(X_1,X_2;Z)$ versus the obtained conditional mutual information $I(X_1;X_2|Z)$.}
        \label{fig:micurve_v2y8}
\end{figure*}
\begin{figure*}
    \centerline{
        \subfloat[$V=3,|\mathcal{Y}|=|\mathcal{Z}|=8$ invertible case]{
            \includegraphics[width=3.0in]{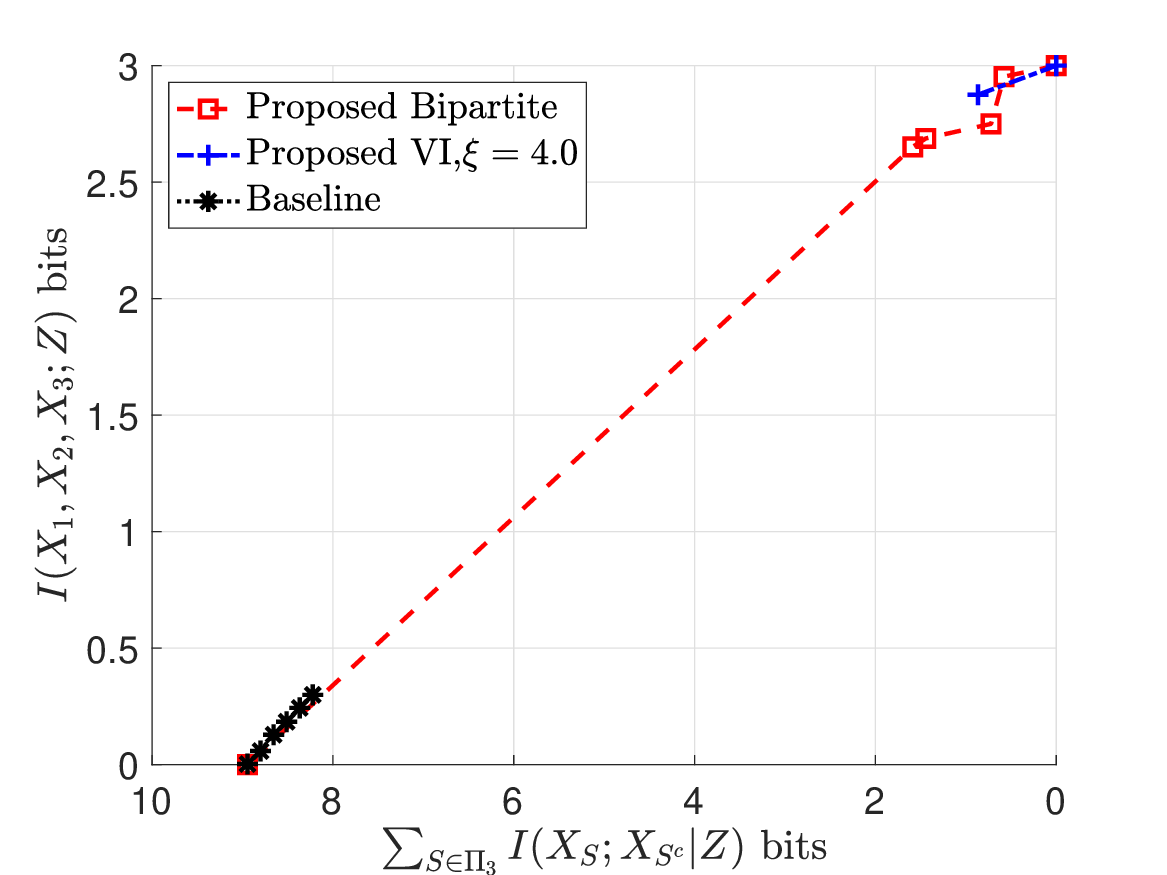}
            \label{subfig:micurve_v3y8c00}
        }
        \hfil
        \subfloat[$V=3,|\mathcal{Y}|=|\mathcal{Z}|=8$ non-invertible case]{
            \includegraphics[width=3.0in]{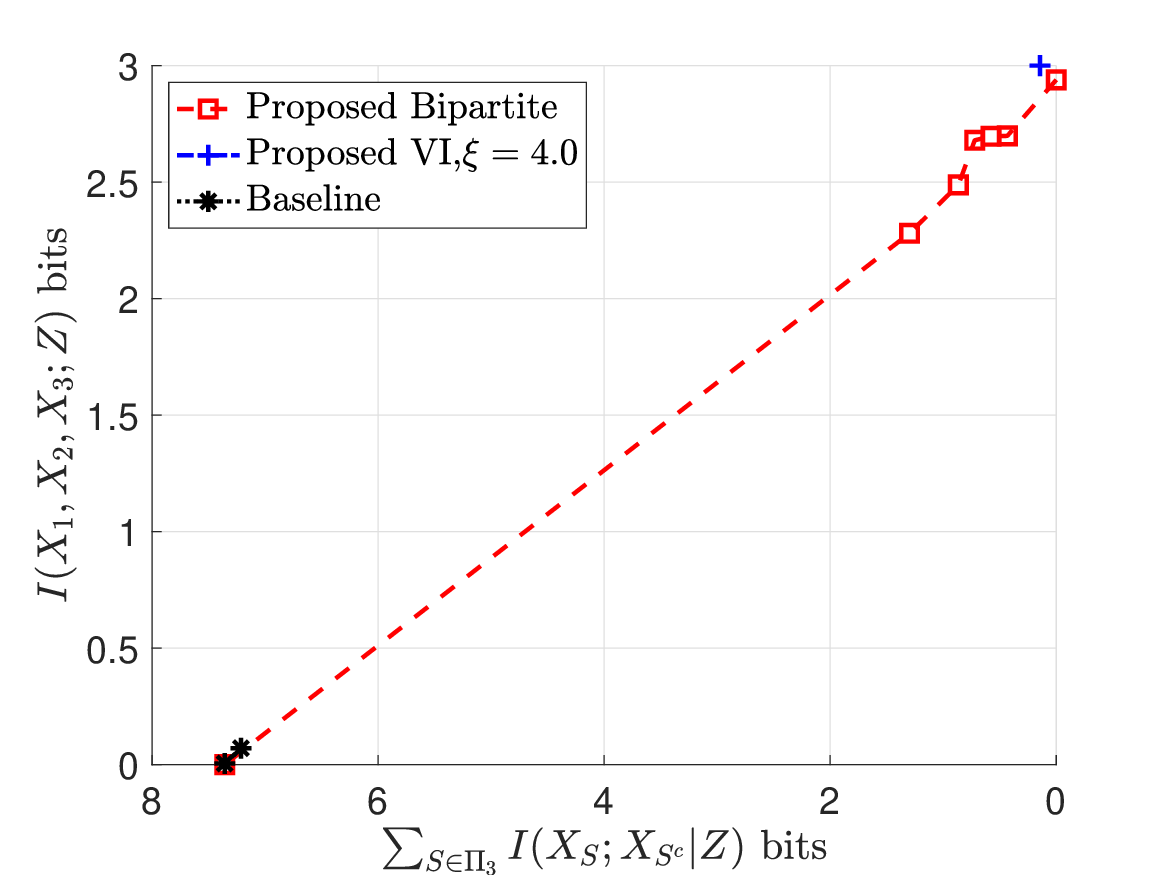}
            \label{subfig:micurve_v3y8c01}
        }}
        \caption{Evaluation results on the information plane. The minimum mutual information $I(X_1,X_2,X_3;Z)$ versus the sum of obtained conditional mutual information $\sum_{i\in\{1,2,3\}}I(X_i;X_{[3]\backslash i}|Z)$.}
        \label{fig:micurve_v3y8}
\end{figure*}

The focus of our evaluation is on the solutions that satisfy Wyner common information conditions, i.e., $P^*_\theta(Z|X^V)$ such that $I(X_i;X_j|Z)\approx 0,\forall i\neq j\in[V]$, with the lowest possible $I(X^V;Z)$. For simplicity, we assume knowing the cardinality of $|\mathcal{Y}|$, so $|\mathcal{Z}|=|\mathcal{Y}|=8$ for all the evaluated solvers. 

In Fig \ref{fig:micurve_v2y8}, we show the lowest achieved mutual information $I(X^V;Z)$ of the solvers versus the obtained sum of conditional mutual information $\sum_{S\in\Pi_V}I(X_S;X_{S^c}|Z)$. This is done by an offline calculation of the obtained solutions as described in the last part. In Fig \ref{subfig:micurve_v2y8c00}, we evaluate the solvers for the invertible case with $V=2$. Both the proposed solvers obtain solutions (almost) satisfying the Wyner common information conditions in contrast to the baseline. Furthermore, among the solutions with $I(X_1;X_2|Z)\approx 0$, both proposed solvers attains comparable $I(Z,X_1,X_2)\approx 3$ bits. It is straightforward to see that it is the optimal solution (in this specific case) under Wyner common information conditions. It is worth noting that our \textit{VI} solver attains the optimal solution with $|\mathcal{Z}||X_1|+|\mathcal{Z}||X_2|=256$ parameters only as compared to $|\mathcal{Z}||X_1||X_2|=2048$ parameters for the other two methods.

Then, we consider the non-invertible case with results shown in Fig. \ref{subfig:micurve_v2y8c01}. Both the proposed solvers again obtain solutions that have $I(X_1;X_2)\approx 0$ contrary to the baseline. However, comparing the two proposed solvers, \textit{Bipartite} achieves better solution ($I(X_1,X_2;Z)\approx 2.6$ bits) than that obtained by \textit{VI}, this implies that \textit{Bipartite} is more robust to problems when there are ambiguity in mapping the observations to the target variables.

Next, we extend the evaluation to $V=3$. Here the evaluation criterion is the lowest possible $I(Z;X_1,X_2,X_3)$ for an obtained sum of conditional mutual information $I(X_1;X_2,X_3|Z)+I(X_2;X_1,X_2|Z)+I(X_3;X_1,X_2|Z)$. The results are shown in Fig. \ref{fig:micurve_v3y8}. For the invertible case (Fig. \ref{subfig:micurve_v3y8c00}), both the proposed solvers again attain the optimal solutions contrary to the baseline. As for the non-invertible case (Fig. \ref{subfig:micurve_v3y8c01}), the \textit{Bipartite} outperforms the other two methods, similar to the insights we have from the $V=2$ settings.

\subsubsection{Clustering Performance}\label{subsubsec:clustering_known}
\begin{figure*}
    \centerline{
        \subfloat[$V=2,|\mathcal{Y}|=|\mathcal{Z}|=8$ invertible case]{
            \includegraphics[width=3.0in]{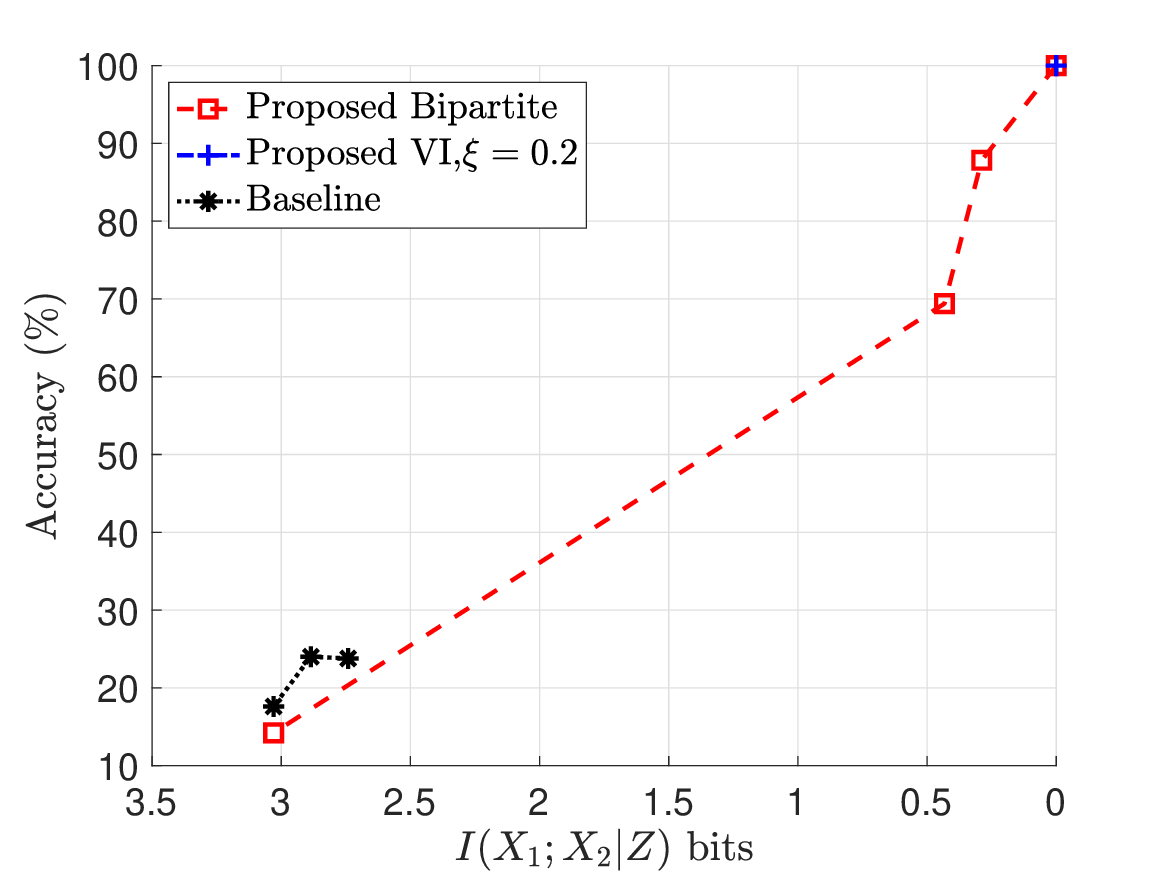}
            \label{subfig:acc_v2y8c00}
        }
        \hfil
        \subfloat[$V=2,|\mathcal{Y}|=|\mathcal{Z}|=8$ non-invertible case]{
            \includegraphics[width=3.0in]{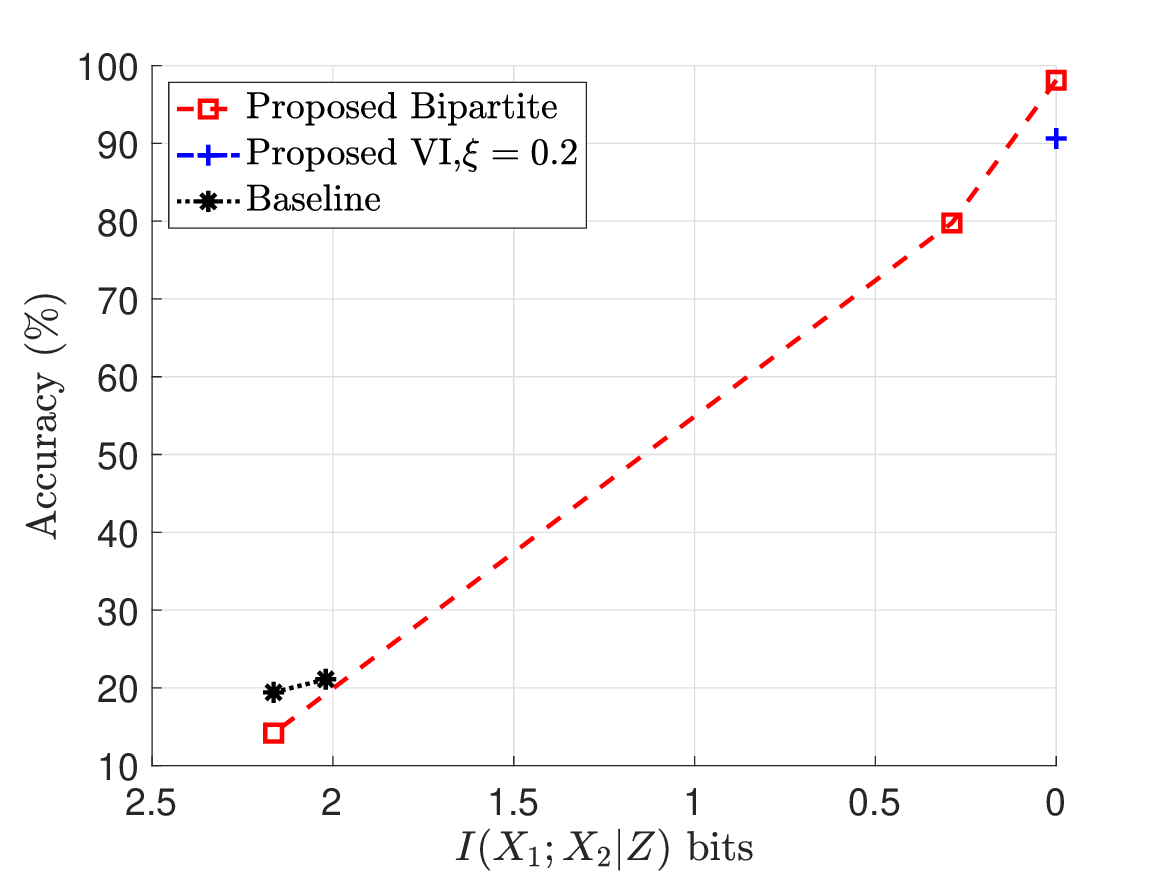}
            \label{subfig:acc_v2y8c01}
        }}
        \caption{Clustering accuracy versus conditional mutual information $I(X_1;X_2|Z)$ of the obtained $P_\theta(Z|X_1,X_2)$ (minimum $I(X_1,X_2;Z)$ achieved) from compared solvers.}
        \label{fig:acc_v2y8}
\end{figure*}
\begin{figure*}
    \centerline{
        \subfloat[$V=3,|\mathcal{Y}|=\mathcal{Z}=8$ invertible case]{
            \includegraphics[width=3.0in]{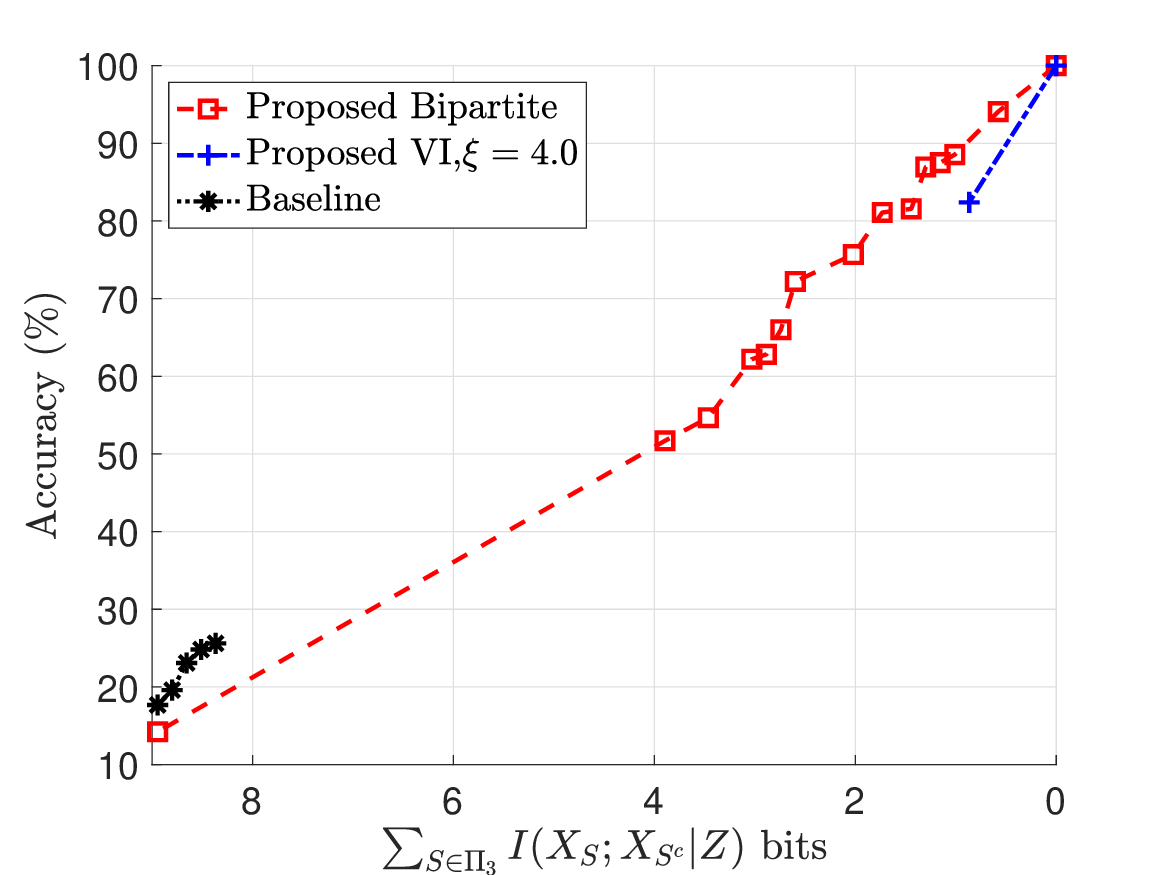}
            \label{subfig:acc_v3y8c00}
        }
        \hfil
        \subfloat[$V=3,|\mathcal{Y}|=|\mathcal{Z}|=8$ noninvertible case]{
            \includegraphics[width=3.0in]{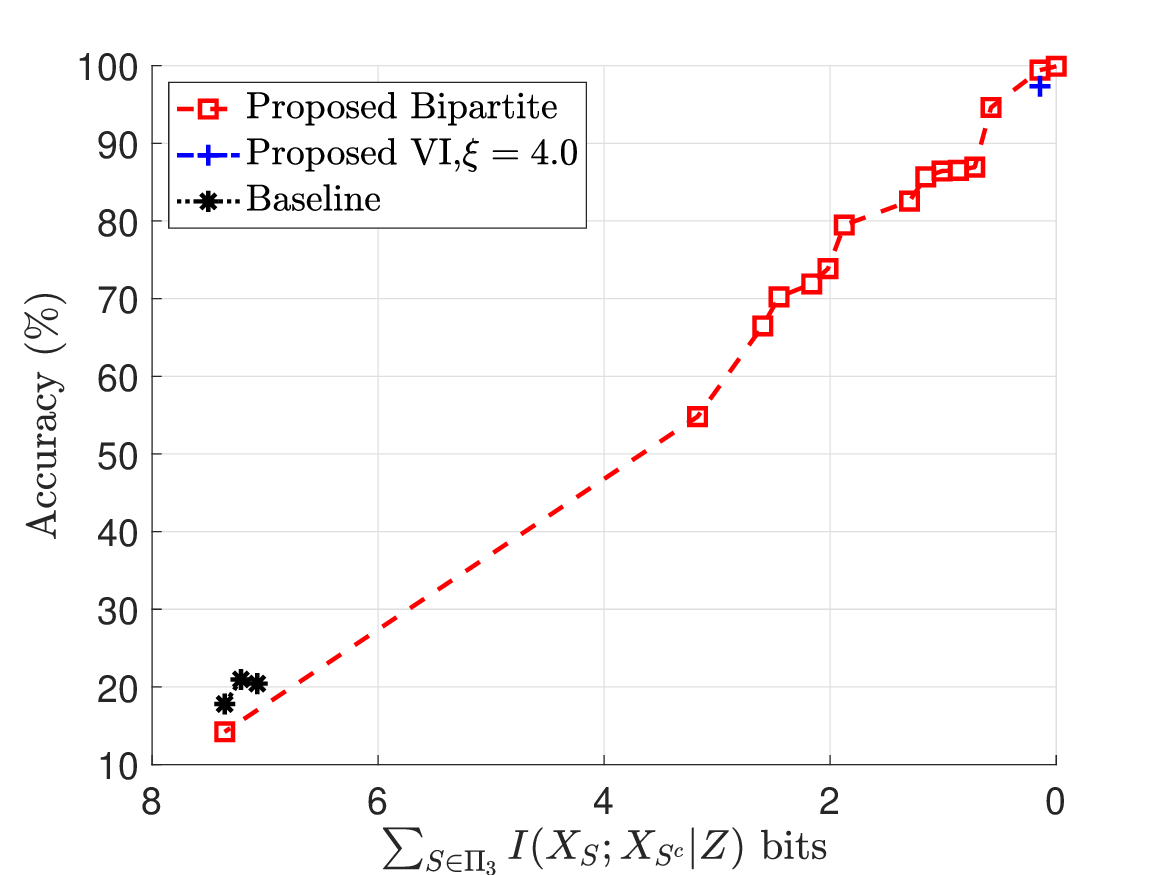}
            \label{subfig:acc_v3y8c01}
        }}
        \caption{Clustering accuracy versus conditional mutual information $\sum_{i\in\{1,2,3\}}I(X_i;X_{[3]\backslash i}|Z)$ of the obtained $P_\theta(Z|X_1,X_2,X_3)$ (minimum $I(X_1,X_2,X_3;Z)$ achieved) from compared solvers.}
        \label{fig:acc_v3y8}
\end{figure*}
We apply the compared solvers to unsupervised multi-modal clustering problems. The common information $Z$ coincides with the cluster target $Y$ (after label matching). This is based on the prior belief that given $Y$ the multiple modalities of observations $X^V$ becomes statistically independent. Moreover, the solution of the solvers is the statistical mappings $P_\theta(Z|X^V)$ from a given $x^V\in\mathcal{X}^V$ to a $z\in\mathcal{Z}$. This can be viewed as a Bayes decoder. The joint distribution $P(X^V)$ does not involve $Y$, which becomes the unlabeled training distribution for a solver. As for testing, a label matching process in clustering literature, e.g., linear sum assignment algorithm~\cite{labelmatching}, is adopted to match ground-truth labels $y\in\mathcal{Y}$ and the estimated $z\in\mathcal{Z}$ for the evaluation of clustering accuracy.

For the testing dataset, we randomly sample the cluster variable $y\in Y$ for $10000$ times, followed by inverse transform sampling~\cite{DevroyeLuc2013NRVG} for all sources $P(X_i|y),\forall i\in[V]$. This gives us a labeled multi-modal dataset $\{(x_1,\cdots,x_V),y\}$ for evaluation. Again, we assume knowing the clustering number $|\mathcal{Z}|=|\mathcal{Y}|$.

For computing the clustering accuracy, we treat the output $P_\theta(Z|X^V)$ of a solver as a Bayes decoder. When given a pair of $x^V$, we first compute the cumulative sum of $P_\theta(Z|x^V)$. Then we get a random sample from a uniformly distributed source $u\in[0,1]$. The index that $u$ is first less than to is the estimated $\hat{y}$. After label matching, we obtain the clustering accuracy.

The results for $V=2$ are shown in Fig. \ref{fig:acc_v2y8}. We evaluate the solvers in both invertible (Fig. \ref{subfig:acc_v2y8c00}) and non-invertible (Fig. \ref{subfig:acc_v2y8c01}) settings. In both cases, the solutions from our solvers approximate conditional independence better, i.e., $I(X_1;X_2|Z)\approx 0$, resulting in the best clustering accuracy. Remarkably, both the proposed solvers attain close-to-optimal solutions in contrast to the baseline. The extended evaluation for $V=3$ is shown in Fig. \ref{fig:acc_v3y8}. Our results clearly demonstrate similar insights (complexity gain of \textit{VI} and robustness of \textit{Bipartite}) from the last part.

\begin{figure*}
    \centerline{
        \subfloat[$V=2$ invertible case]{
            \includegraphics[width=3.0in]{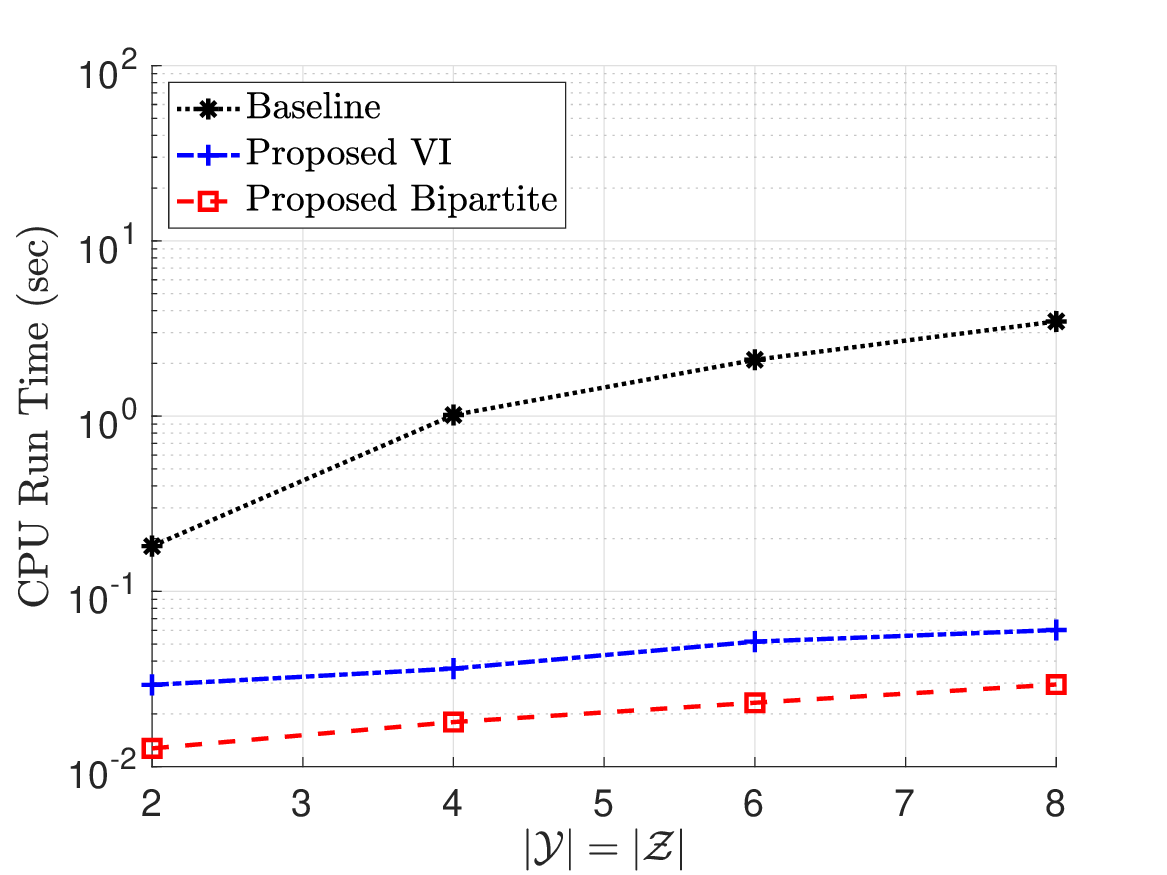}
            \label{subfig:niter_v2corr0}
        }
        \hfil
        \subfloat[$V=3$ non-invertible case]{
            \includegraphics[width=3.0in]{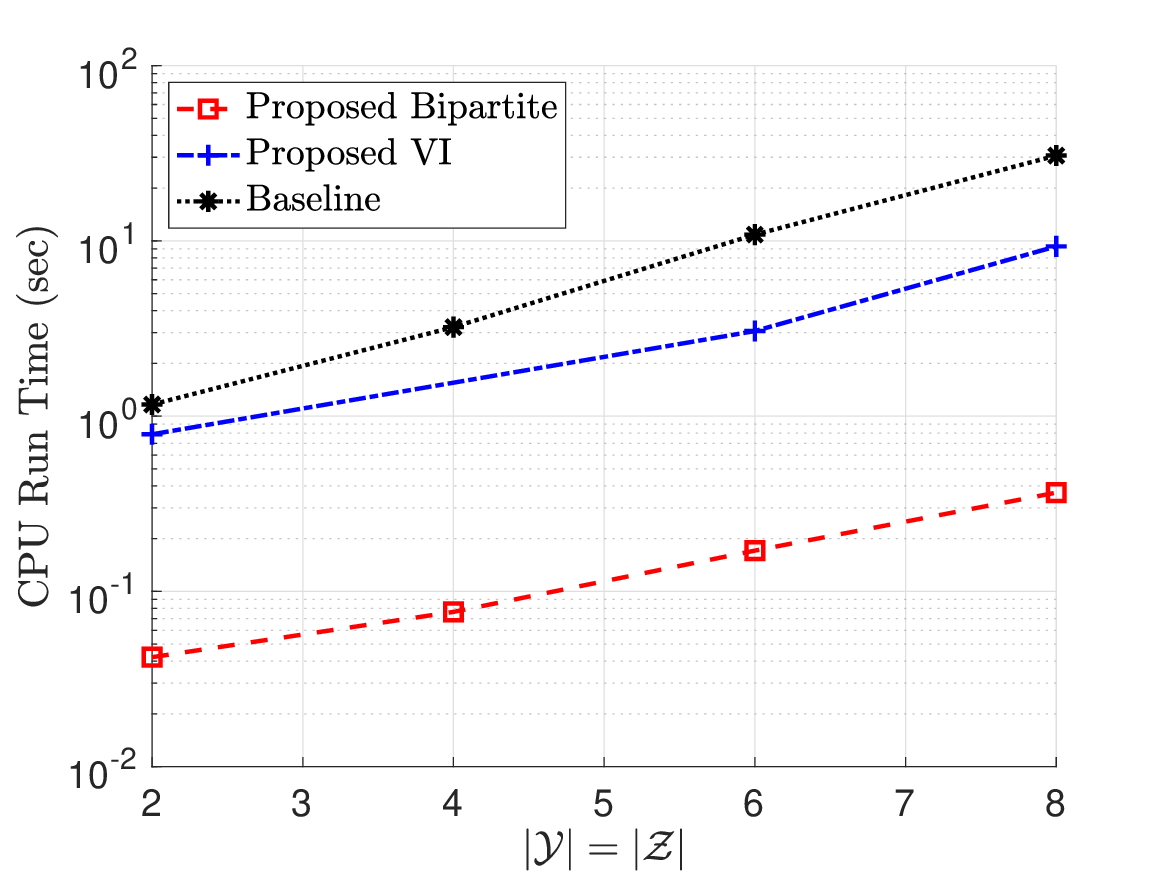}
            \label{subfig:niter_v3corr0}
        }
    }
    \caption{Running time performance of the compared solvers. Assume knowing the number of cardinality $|\mathcal{Y}|$, hence setting $|\mathcal{Z}|=|\mathcal{Y}|$.}
    \label{fig:niter_v23}
\end{figure*}

Finally, we compare the running time performance of the compared solvers. We run all solver on the same commercial laptop and record the time from initialization to the completion of the procedures mentioned in the end of Sec. \ref{subsubsec:cmp_methods}. We vary the cardinality $|\mathcal{Z}|$ and report the running time. The results are shown in Fig. \ref{fig:niter_v23}. Both the proposed solvers finish the experiments significantly faster than the baseline, with maximum $2$ order less running time required. This clearly demonstrates the benefit of our closed-form update equations~\eqref{eq:wyneram_nv_main} and \eqref{eq:wynerdca}. Comparing the proposed solvers, \textit{VI} requires more running time. We empirically find that the \textit{VI} solver is more dependent on an initialization point, hence a patience-restart mechanism is implemented in our first prototype which is triggered occasionally. A more optimized prototype is left for future work.


\subsection{Evaluation for Unknown Joint Distribution}\label{subsec:eval_unknown}
\subsubsection{Datasets}
\begin{table*}[!t]
\begin{minipage}{\textwidth}
\centering
\caption{Details of the Evaluated Datasets}
\label{table:datasets}
\renewcommand{\arraystretch}{1.3}
\begin{threeparttable}
\renewcommand*{\thempfootnote}{\fnsymbol{mpfootnote}}
\begin{tabular}{ccccc}
\hline\hline
Datasets & Number of Samples (Train/Test) & Number of Modalities & Number of Clusters & Dimensions (Flattened) \\ \hline
MNIST-USPS & 4500/500 & 2 & 10 & 784, 784 \\ \hline
Fashion & 9000/1000 & 3 & 10 & 784, 784, 784 \\ \hline
BDGP & 2250/250 & 2 & 5 & 1750, 79\\ \hline
Caltech-3V & 1260/140 & 2 & 7 & 40, 254, 928 \\ \hline
Caltech-4V & 1260/140 & 3 & 7 & 40, 254, 928, 512\\ \hline
Caltech-5V & 1260/140 & 4 & 7 & 40, 254, 928, 512, 1984\\ \hline
CCV & 6084/689 & 3 & 20 & 5000, 5000, 4000\\ \hline
PolyMNIST & 60000/10000 & 5 & 10 & 784, 784, 784, 784, 784\\ \hline
\end{tabular}
\begin{tablenotes}
    \footnotesize
\end{tablenotes}
\end{threeparttable}
\end{minipage}    
\end{table*}
In this part we focus on multi-modal clustering problems where empirical samples of the multi-modal observations $X^V$ (training set) are available instead of the joint distribution. Moreover, the tarining set is given without ground-truth cluster labels. The evaluation includes following recent datasets~\cite{xu2022MFLVC}: 
\begin{itemize}
    \item MNIST-USPS~\cite{dtMnistUsps}: The dataset consists of two well-known datasets in computer vision communities: MNIST~\cite{mnistdataset} and USPS~\cite{uspsdtorig}. Both are images of digits ($10$ clusters). There are many post-processed versions in literature. We use the one provided in~\cite{xu2022MFLVC}.
    \item Fashion~\cite{xu2021multivae}: Modified from~\cite{dtfashion}. The original dataset consists of gray-scale images of $10$ types of clothing (single-modal). In~\cite{xu2021multivae}, three products of the same category are paired as a multi-modal sample.
    \item Berkeley Drosophila Genome Project (BDGP)~\cite{dtBDGP}: Patches of gray-scale images of \textit{Drosophila} embryos are processed into three features: lateral, dorsal and ventral. The features vectors are concatenated together as the first modality. The corresponding control terms are the second modality, paired with the first one. 
    \item Caltech~\cite{dtCaltech}: \cite{li2015large} selected $7$ categories from the original colored images dataset~\cite{dtCaltech} of common objects. A variety of features of the images is extracted as different modalities.~\cite{xu2022MFLVC} selected up to five modalities: wavelet moment, CENTRIST~\cite{dttypecentrist}, LBP~\cite{dttypelbp}, GIST~\cite{dttypegist} and HoG~\cite{dttypehog}.
    \item Columbia Consumer Video (CCV)~\cite{dtCCV}: Three visual/audio features are extracted from the videos of the dataset as different modalities. The features are standard: Static SIFT~\cite{dttypesift}, Spatial-temporal interest points (STIP)~\cite{dttypestip} and Mel-frequency cepstral coefficients (MFCC).
    \item PolyMNIST~\cite{sutter2021generalized,hwang2021multi,palumbo2023mmvae+}: Adapted from the MNIST hand-written digits dataset (images), where a sample is one of the 10 digits written in difference styles, each synthesized with a different background.
\end{itemize}
The above datasets have various modalities, dimensions of features and sample size. We summarize these attributes in Table \ref{table:datasets}. The details is referred to~\cite[Sec. 4]{xu2022MFLVC} and~\cite{sutter2021generalized}.

In this part, we assume knowing the number of clusters. For the training phase, a dataset without labels is provided to each solver whereas in the testing phase, the computed cluster labels of each sample of the same dataset are compared with the corresponding ground-truth label to obtain accuracy rate. This is done after labels matching~\cite{labelmatching,xu2021multivae,Conan21,asano2020labelling,alwassel2020self} as described in Section \ref{subsubsec:clustering_known}.

\subsubsection{Compared Methods}
Here, our approaches are compared with the following state-of-the-art methods. We summarize the novelty of ours as compared to each of them.
\begin{enumerate}[i)]
    \item MFLVC~\cite{xu2022MFLVC}: The method has the closest model architecture as ours. However, the multi-modal low-dimensional features are subjected to the contrastive multiview coding~\cite{tian2020contrastive} loss as a heuristic sub-objective whereas ours follows by Theorem \ref{thm:multimodal_mi_up_bound}. For clustering, the method parameterizes each modality with categorical distribution as well. However, ad-hoc KMeans clustering algorithm is applied to each extracted modal-specific feature and optimizes the categorical distribution with self-supervised learning. Our approaches do not need any of these heuristics. Finally, the final prediction of this method is a simple average of all model-specific predictions, whereas ours is based on the insight from the self-consistent equation~\eqref{eq:impl_categorical_qprob}. 
    
    \item PoE~\cite{poe2020feifeiPoE}: This method implements variational autoencoders (VAE)~\cite{kingma2013auto} for each modality, and a common Gaussian random variable is formed similar to \eqref{eq:impl_gaussian_poe} assuming the covariance matrices are diagonal. Then a Gaussian common feature is sampled from the equivalent mean and covariance matrices. This common feature is given to each modal-specific decoder to reconstruct the multi-modal observations. The method does not apply to multi-modal clustering directly, so we add an extra KMeans algorithm to cluster the Gaussian common features. In contrast, our methods can be directly applied to clustering problems by the categorical parameterization. Additionally, Theorem \ref{thm:multimodal_mi_up_bound} applies to our methods which leverages the fact that multi-modal samples come in pairs different from this method.
    \item MVTCAE~\cite{hwang2021multi}: This method maximizes the total correlation objective: $\sum_{i=1}^VI(Z;X_i)-I(Z;X^V)$ which is equivalent to the special case of the \textit{Variational} formulation with $\kappa_{[V]}=1$ without the correlation maximization step. Different from this method, our methods adopt Theorem \ref{thm:multimodal_mi_up_bound} to leverage the fact that the multi-modal samples come in pairs. In our evaluation, we modify it by employing a KMeans clustering of the cascaded representations. 
    \item MoE~\cite{shi2019variationalMOE,palumbo2023mmvae+}: This method implements VAE for each modality as well. However, the Gaussian sample from one modal-specific encoder is passed to all decoders to reconstruct the observations for all modalities. We concatenate all modal-specific Gaussian feature vectors into a long vector and add an extra KMeans algorithm to cluster it. The method focuses on estimating the joint distribution by the generalized evident lower bound~\cite{sutter2021generalized} whereas we bypass it and extract the common randomness for clustering. 
    \item MoPoE~\cite{sutter2021generalized}: The method also implements VAEs as data processing modules for dimension reductions of each modality of observations. The loss function for minimization of this method combines both PoE and MoE, hence its name. For clustering problems, the low-dimensional feature vectors are concatenated together as a long vector then applied to a KMeans algorithm. The details are referred to~\cite{sutter2021generalized}. The differences to our methods are as mentioned previously.
\end{enumerate}

In summary, our solvers stand out from the compared methods in the following senses. Contrary to the MFLVC baseline that consists of several heuristics, our solvers are based on the proposed common information frameworks~\eqref{eq:unknown_bipartite_prob},\eqref{eq:vi_reduction_unknow}, with Theorem \ref{thm:multimodal_mi_up_bound} applied. 
As compared to PoE, MVTCAE, MoE and MoPoE: our method is based on the difference-of-convex structure of the \textit{Bipartite} and \textit{Variational} frameworks instead of the generalized evident lower bound. The fundamental difference is that our clustering solver is based on direct parameterization of the common information to avoid estimating the joint distribution $P(X^V)$, whereas these baselines essentially estimate a lower bound of the log-likelihoods of the joint distribution $\log P(X^V)\geq \mathbb{E}[P(Z)P(X^V|Z)/Q_\theta(Z|X^V)]$~\cite{sutter2021generalized}.

We are aware of other approaches such as multi-view spectral clustering~\cite{kumar2011co,spectralclustering2019,SC2018rankfact}. Similar to standard spectral clustering, the multi-view extension relies on spectral decomposition to extract common features but then the view-specific features are joint used for clustering. However, due to the low-rank matrix factorization problem involves and the high dependency of the similarity matrix required, these methods suffer from higher computation complexity than the listed compared methods. Therefore, we leave the comparison to spectral clustering methods as future works.

For the baselines, we use their latest prototypes available online with minor modification on the network architecture for a fair comparison\footnote{MFLVC baseline is available at https://github.com/SubmissionsIn/MFLVC; MVTCAE is available at: https://github.com/gr8joo/MVTCAE with MoPoE, MoE and PoE prototypes included. We are aware of recent methods, e.g., MMVAE++ (available at: https://openreview.net/forum?id=sdQGxouELX), but after modification their clustering performance are worse than the presented representative method in our evaluation, and hence are omitted.}. All methods are run on a commercial computer with the same experiment settings. For each experiment, we randomly initialize the parameters with a random seed, train the model with the designed objective function for $300$ iterations and output the parameters reaching the minimum training loss, we repeat the procedures for $10$ different random seeds and report the average clustering accuracy. The default hyperparameters and optimizers of the available prototypes are adopted.

\begin{table*}[!t]
\begin{minipage}{\textwidth}
\centering
\caption{Accuracy of Evaluated Methods and Datasets. Scale of $[0,1]$}
\label{table:eval_acc}
\renewcommand{\arraystretch}{1.3}
\begin{threeparttable}
\renewcommand*{\thempfootnote}{\fnsymbol{mpfootnote}}
\begin{tabular}{cccccccc}
\hline
 & Bipartite (Ours) & VI (Ours) & MFLVC & MVTCAE& MoPoE & MoE & PoE \\ \hline\hline
 BDGP & \textbf{0.97} & 0.96 & 0.92 & 0.25 &0.24 & 0.24 & 0.24 \\ \hline
Caltech-3V & \textbf{0.66} & 0.65 & 0.64 & 0.60 & 0.63 & 0.59 & 0.65 \\ \hline
Caltech-4V & \textbf{0.73} & \textbf{0.73} & 0.70 & 0.53 & 0.71 & 0.71 & 0.71 \\ \hline
Caltech-5V & \textbf{0.79} & 0.76 & 0.77 & 0.58 &0.62 & 0.69 & 0.63 \\ \hline
CCV & 0.28 & \textbf{0.29} & 0.28 & 0.20 & 0.20 &0.21 & 0.20 \\ \hline
Fashion & \textbf{0.99} & \textbf{0.99} & 0.98 & 0.90& 0.67 & 0.50 & 0.67 \\ \hline
MNIST-USPS & \textbf{0.99} & \textbf{0.99} & \textbf{0.99} & 0.50& 0.64 & 0.71 & 0.52 \\ \hline
PolyMNIST & \textbf{0.99} & \textbf{0.99} & \textbf{0.99} & 0.84 & 0.71 & 0.22 & 0.62  \\ \hline
\end{tabular}
\end{threeparttable}
\end{minipage}
\end{table*}

\begin{table*}[!t]
\begin{minipage}{\textwidth}
\centering
\caption{Running Time (Training) For $300$ Epochs. In Seconds}
\label{table:eval_tr_time}
\renewcommand{\arraystretch}{1.3}
\begin{threeparttable}
\renewcommand*{\thempfootnote}{\fnsymbol{mpfootnote}}
\begin{tabular}{cccccccc}
\hline
 & Bipartite (Ours) & VI (Ours) & MFLVC& MVTCAE & MoPoE & MoE & PoE \\ \hline\hline
BDGP & \textbf{405.58} & {410.72} & 415.65& 465.21 & 487.40 & 470.39 & 494.45 \\ \hline
Caltech-3V & 435.01 & \textbf{402.05} & 405.55 & 442.94 & 456.19 & 449.64 & 463.94 \\ \hline
Caltech-4V & 703.34 & \textbf{414.69} & 419.15 & 447.65& 470.11 & 456.17 & 480.67 \\ \hline
Caltech-5V & 1662.42 & \textbf{430.89} & 436.92 & 478.19& 517.80 & 474.11 & 531.49 \\ \hline
CCV & 820.66 & \textbf{595.52} & 632.64 & 643.02& 702.08 & 650.40 & 767.29 \\ \hline
Fashion & 966.71 & \textbf{628.14} & 663.47 & 667.65& 785.25 & 674.83 & 935.86 \\ \hline
MNIST-USPS & \textbf{445.61} & {451.07} & 462.90 & 519.07 & 555.80 & 524.93 & 579.47 \\ \hline
PolyMNIST & 4045.69 & \textbf{3721.40} & 3804.80 & 4387.87& 5897.52& 3855.36 & 5513.83  \\ \hline
\end{tabular}
\end{threeparttable}
\end{minipage}
\end{table*}

\subsubsection{Neural network-based prototypes}\label{subsubsec:nn_impl}
We implement the proposed \textit{Bipartite} and \textit{Variational} solvers with neural networks. The computation flows are as shown in Algorithm \ref{alg:pseudo_bipartite} and \ref{alg:pseudo_wvae}. Here, we provide details of the network architectures.

For the datasets: BDGP, Caltech, CCV, Fashion, MNIST-USPS, we follow the autoencoders adopted in the MFLVC baseline~\cite{xu2022MFLVC} (the state-of-the-art of the listed datasets). The encoder consists of $4$ linear layers with standard Rectified Linear Unit (ReLU) activation employed except for the encoder output. The number of neurons for each layer is: $d^{(v)}-500-500-2000-512$, where $d^{(v)}$ is the number of dimensions of the $v$-th modality. The decoder is simply a reverse of the above configurations. For the \textit{Bipartite} solver, we implement a set of single-layer linear weights of $|W_S|=512\times |S|$ input dimensions and $128$ output dimensions for the correlation maximization step, the similarity function follows the one adopted in~\cite{xu2022MFLVC,tian2020contrastive}. For the parameterized categorical distributions, we add another set of single-layer linear weights with Softmax activation of $|W_S|$ input dimensions and $|\mathcal{Z}|$ output dimensions. As for the \textit{Variational} solver, a single linear transform of $512\times 128$ weights is shared between all $W_i,\forall i\in[V]$ encoder outputs for correlation maximization and similarly for the parameterized distributions where a $512\times |\mathcal{Z}|$ linear transform is shared for all encoder outputs.

As for the PolyMNIST dataset, we follow the autoencoder architectures given in~\cite{hwang2021multi}. The architecture is a standard convolutional neural network (CNN). The encoder consists of $3$ stacks of two-dimensional convolutional layers with kernel size $3$, stride $2$, padding $1$ and $32-64-128$ output channels. Then the output of the CNN stacks is flattened and fed to a linear layer of $2048\times 512$ weights with ReLU activation. The decoder reverses the configurations. The architectures of other modules follow what described in the last paragraph. 

In optimizing the neural networks, we use the standard ADAM optimizer with a learning rate $3\times 10^{-4}$~\cite{kingma2014adam}. The other hyperparameters align with the two available source codes~\cite{xu2022MFLVC,hwang2021multi}.

For the hardware, we implement the proposed solvers in PyTorch environment and run on a commercial computer with $8$GB memory with Nvidia-V100 graphic card support.

\subsubsection{Clustering Performance}
For a fair comparison, all compared methods implement autoencoders as the dimension reduction step. We align the architectures of the autoencoders for all compared methods as described in Section \ref{subsubsec:nn_impl}. Therefore, our evaluation can be interpreted as comparing the relevance of the extracted common features for multi-modal clustering. In other words, we evaluate the effectiveness of the proposed loss functions when the target task is multi-modal clustering. The evaluation metrics is the clustering accuracy, computed after label matching.

The results are shown in Table \ref{table:eval_acc}. The highest accuracy achieved among the evaluated methods is highlighted in boldface. In most cases, the proposed \textit{Bipartite} and \textit{Variational} solvers obtain higher accuracy over all compared baselines. As for other cases where the proposed solver attain comparable performance (accuracy $\approx 1.0$) to some of the baselines, our solvers are more efficient than them with significantly reduced training time. The training time performance is shown in Table \ref{table:eval_tr_time} whereas the inference phase running time is shown in Table \ref{table:eval_ts_time}. In particular, the proposed \textit{Bipartite} requires $3.7\%$ less running time than the second-best approach (versus MFLVC for MNIST-USPS dataset).

Overall, the evaluation demonstrates that the proposed solvers have improved performance in the evaluated multi-modal clustering problems than the state-of-the-art. We achieve higher accuracy than the compared approaches; For methods that have comparable accuracy to ours, we still outperform them for less running time. This clearly demonstrates the advantages of the proposed Wyner frameworks.

\begin{table*}[!t]
\begin{minipage}{\textwidth}
\centering
\caption{Running Time (Inference) of Testing Datasets. In Seconds}
\label{table:eval_ts_time}
\renewcommand{\arraystretch}{1.3}
\begin{threeparttable}
\renewcommand*{\thempfootnote}{\fnsymbol{mpfootnote}}
\begin{tabular}{cccccccc}
\hline
 & Bipartite (Ours) & VI (Ours) & MFLVC & MVTCAE & MoPoE & MoE & PoE \\ \hline\hline
BDGP & \textbf{0.016} & \textbf{0.016} & 0.018& 1.60 & 1.58 & 1.45 & 1.56 \\ \hline
Caltech-3V & 0.020& \textbf{0.014} & \textbf{0.014}& 1.56 & 1.59 & 1.50 & 1.60 \\ \hline
Caltech-4V & {0.024}& 0.018 & \textbf{0.015} & 1.58& 1.55 & 1.55 & 1.56 \\ \hline
Caltech-5V & 0.035 & \textbf{0.019} & 0.023 & 1.50 & 1.60 & 1.57 & 1.61 \\ \hline
CCV & {0.065}& \textbf{0.051} & 0.062 & 1.58& 1.62 & 1.51 & 1.56 \\ \hline
Fashion & {0.078}& \textbf{0.066} & 0.132 & 1.54& 1.56 & 1.57 & 1.61 \\ \hline
MNIST-USPS & \textbf{0.028} & \textbf{0.028} & 0.044 & 1.49& 1.54 & 1.43 & 1.55 \\ \hline
PolyMNIST & 1.76 & 1.85 & \textbf{1.64} & 2.10 & 2.00 & 1.90 & 1.83  \\ \hline
\end{tabular}
\end{threeparttable}
\end{minipage}
\end{table*}


\section{Conclusions}
We study two novel extensions of the Wyner common information. For \textit{Variational} common information, we relax the matching condition and focus on feasible solution set that satisfies conditional independence. We propose a more efficient and simplified alternating minimization solver based on this insight. As for \textit{Bipartite} common information, we instead relax the conditional independence which reveals the difference-of-convex structure of the problem. Inspired by DCA, we propose a new efficient solver based on the derived closed-form self-consistent equations. We prove the convergence (local minimum) of the proposed solvers, in contrast to previous solvers. Then we generalize the proposed solvers to unknown distributions settings with data samples available. By employing the mixture of members of the exponential family, the parameters of the proposed solvers can be optimized with efficient MLE. In application, we show that the proposed Wyner frameworks and solvers are a good fit for unsupervised multi-modal clustering problems. Empirically, the proposed solvers outperform the state-of-the-art in terms of both clustering accuracy and running time. For future work, we aim to study the sample complexity of the proposed solvers in terms of the number of samples required for convergence.

\bibliographystyle{IEEEtran}
\bibliography{references}


%

\appendices


\section{Proof of Lemma \ref{lemma:vi_full_know_ub}}\label{appendix:pf_lemma_vi_full_know_ub}
The derivations follow by:

\begin{IEEEeqnarray}{rCl}
    I_\theta(X^V;Z)&=&\mathbb{E}_Z\left[\sum_{x^V\in\mathcal{X}^V}P_\theta(x^V|Z)\log{\frac{P_\theta(x^V|Z)}{\sum_{z'\in\mathcal{Z}}P(z')P_\theta(x^V|z')}}\right]\nonumber\\
    &=&-H_\theta(X^V|Z)-\mathbb{E}_{X^V;\theta}\left[\log{\sum_{z'\in\mathcal{Z}}P(z')P_\theta(X^V|z')\frac{P(X^V)}{P(X^V)}}\right]\nonumber\\
    &=&-\sum_{i=1}^VH_\theta(X_i|Z)-D_{KL}[P_\theta(X^V)\parallel P(X^V)]-\sum_{x^V\in\mathcal{X}^V}\left[\sum_{z\in\mathcal{Z}}P(z)\prod_{i=1}^VP_\theta(x_i|z)\right]\log{P(x^V)}\nonumber\\
    &\leq&-\sum_{i=1}^VH_\theta(X_i|Z)-\mathbb{E}_{X^V;\theta}\left[\log{P(X^V)}\right],\IEEEeqnarraynumspace\IEEEyesnumber\label{eq:var_dkl_full_know_mi_bound}
\end{IEEEeqnarray}
where the last equality is due to the construction $P(X^V|Z)=\prod_{i=1}^VP(X_i|Z)$ and the inequality is due to the non-negativity of KL divergence.

\section{Derivation for the VI Solver}\label{appendix:wyner_am_derivation}
Without loss of generality, consider the update for the first modality, i.e., $P(X_1|Z)$. And for simplicity, let us start with the case $V=2$. The part of the objective function that it depends on is given by:
\begin{IEEEeqnarray}{rCl}
\mathcal{L}_1:=-H(X_1|Z)-\mathbb{E}_{X_1,X_2;\theta}\left[\log{P(X_1,X_2)}\right]+\beta D_{KL}[P_\theta(X_1,X_2)\parallel P(X_1,X_2)].\IEEEeqnarraynumspace\IEEEyesnumber
\end{IEEEeqnarray}
Denote $P_\theta(X_1,X_2)=\sum_{z\in\mathcal{Z}}p(z)P(X_1|z)P(X_2|z)$, from elementary functional derivative analysis, we have:
\begin{IEEEeqnarray}{rCl}
    -\frac{\partial H_\theta(X_1|Z)}{\partial p_\theta(x_1|z)}&=&p(z)\left[\log{p_\theta(x_1|z)}+1\right],\IEEEyesnumber\IEEEeqnarraynumspace\IEEEyessubnumber\\
    \frac{\partial p_\theta(x_1,x_2)}{\partial p_\theta(x_1|z)}&=&p(z)p_\theta(x_2|z),\IEEEeqnarraynumspace\IEEEyessubnumber\\
    \frac{\partial D_{KL}[p_\theta(x_1,x_2)\parallel p(x_1,x_2)]}{\partial p_\theta(x_1|z)}&=&\sum_{x_2\in\mathcal{X}_2}p(z)p_\theta(x_2|z)\left[\log{\frac{p_\theta(x_1,x_2)}{p(x_1,x_2)}}+1\right].\IEEEeqnarraynumspace\IEEEyessubnumber
\end{IEEEeqnarray}
Therefore, we have:
\begin{IEEEeqnarray}{rCl}
    \frac{\partial\mathcal{L}_1}{\partial p_\theta(x_1|z)}=p(z)\left[\log p_\theta(x_1|z)+\sum_{x_2\in\mathcal{X}_2}p_\theta(x_2|z)\left(-\log{p(x_1,x_2)}+\beta\log{\frac{p_\theta(x_1,x_2)}{p(x_1,x_2)}}\right)+C_z\right],\IEEEeqnarraynumspace\IEEEyesnumber\label{eq:fonc_wyneram}
\end{IEEEeqnarray}
where $C_z$ is a constant. Letting \eqref{eq:fonc_wyneram} equal to zero, with some arrangements, we get:
\begin{IEEEeqnarray}{rCl}
    \log{p_\theta(x_1|z)}=\log{\frac{p(x_1)^{1+\beta}}{p_\theta(x_1)^{\beta}}}+\sum_{x_2\in\mathcal{X}_2}p_\theta(x_2|z)\left[-(1+\beta)\log{\frac{p_\theta(x_2
|z)}{p(x_2|x_1)}}+\beta\log{\frac{p_\theta(x_2|z)}{p_\theta(x_2|x_1)}}\right]+C',\IEEEeqnarraynumspace\IEEEyesnumber\label{eq:wyneram_log2v}
\end{IEEEeqnarray}
for some constant $C'$. Taking exponential operator at both sides of \eqref{eq:wyneram_log2v}, we have:
\begin{IEEEeqnarray}{rCl}
    p_\theta(x_1|z)=\frac{p(x_1)}{M(z)}\left(\frac{p(x_1)}{p_\theta(x_1)}\right)^{\beta}\exp\left\{-(1+\beta)D_{KL}[p_\theta(x_2|z)\parallel p(x_2|x_1)]+\beta D_{KL}[p_\theta(x_2|z)\parallel p_\theta(x_2|x_1)]\right\},\IEEEeqnarraynumspace\IEEEyesnumber\label{eq:wyneram_dkl_diff}
\end{IEEEeqnarray}
where $M(z)$ is a normalization constant for $p_\theta(x_1|z)$ to be a valid conditional probability, and we define $p_\theta(x_1):=\sum_{x_2\in\mathcal{X}_2}p_\theta(x_1,x_2)$, $p_\theta(x_2|x_1):=p_\theta(x_1,x_2)/p_\theta(x_1)$. Equivalently, we can express the above as:
\begin{IEEEeqnarray}{rCl}
    p_\theta(x_1|z)=\frac{p(x_1)}{M(z)}\exp\left\{-D_{KL}[p_\theta(x_2|z)\parallel p(x_2|x_1)]+\beta \sum_{x_2\in\mathcal{X}_2}p_\theta(x_2|z)\log{\frac{p(x_2,x_1)}{p_\theta(x_2,x_1)}}\right\}.\IEEEeqnarraynumspace\IEEEyesnumber
\end{IEEEeqnarray}
So, the trade-off between fitting $P(X_1,X_2)$ and the tolerance of approximation error to $P_\theta(X_1,X_2)$ is revealed. One can simply generalize the above results to arbitrary $V$ by applying the conditional independence $P_\theta(X^V|Z)=\prod_{i=1}^VP_\theta(X_i|Z)$. Denote $[W]:=[V]\backslash \{1\}$:
\begin{IEEEeqnarray}{rCl}
    p_\theta(x_1|z)=\frac{p(x^w)}{M_w(z)}\exp\left\{-D_{KL}[p_\theta(x^w|z)\parallel p(x^w|x_1)]+\beta\sum_{x^w\in\mathcal{X}^W}p_\theta(x^w|z)\log{\frac{p(x_1,x^w)}{p_\theta(x_1,x^w)}}\right\},\IEEEeqnarraynumspace\IEEEyesnumber
\end{IEEEeqnarray}
where $x^w:=(x_2,\cdots,x_V)$ and $X^W:=(X_2,\cdots,X_V)$ similarly.

and hence we complete the proof.


\section{Derivation for the Bipartite Solver}\label{appendix:wyner_dca_derivation}
By our construction, the sub-objective functions $f(P_{z|x^V}),g(P_{z|x^V})$ are defined as in \eqref{eq:dca_fg}. Then the functional derivatives that we need are:
\begin{IEEEeqnarray}{rCl}
    \frac{\partial H(z|x^V)}{\partial p(z|x^V)}:&=&-p(x^V)\left[\log{p(z|x^v)}+1\right],\\
    \frac{\partial H(z)}{\partial p(z)}:&=&p(x^V)\left[\log{p(z)}+1\right],\\
    \frac{\partial I(z;x^W)}{\partial p(z|x^V)}:&=& \frac{\partial \left[H(z)-H(z|x^W)\right]}{\partial p(z|x^V)}=p(x^V)\left[\log{\frac{p(z|x^W)}{p(z)}}\right].
\end{IEEEeqnarray}
Define denote the linearized $g$ as $\bar{g}(P_{z|x^V})$. Also, let $\mathcal{L}^k:=f(P_{z|x^V})-\bar{g}(P^k_{z|x^V})$ with a step $k$ solution. Then the first-order functional derivative of the linearized sub-problem is given by:
\begin{IEEEeqnarray}{rCl}
    \frac{\partial \mathcal{L}^k}{\partial p(z|x^v)}:=p(x^V)\left\{\log{p(z|x^V)}-\log{p^k(z)}-\sum_{S\in\Pi_V}\kappa_S\left[\log{\frac{p(x_S|z)}{p(x_S)}}+\log{\frac{p(x_{S^c}|z)}{p(x^{S^c})}}\right]\right\}\IEEEeqnarraynumspace\IEEEyesnumber\label{eq:appendix_dca_fonc}.
\end{IEEEeqnarray}
Letting \eqref{eq:appendix_dca_fonc} to be equal to zero, followed by some arrangements we get:
\begin{IEEEeqnarray}{rCl}
    p^*(z|x^V):= \frac{p^k(z)}{M^k(\{\kappa_S\},x^V)}\exp\left\{\sum_{S\in\Pi_V}\kappa_S\left[\log{p^k(x_S|z)}+\log{p^k_{x_{S^c}|z}}\right]\right\},
\end{IEEEeqnarray}
where $M^k(\{\kappa_S\},x^V)$ denotes the step-$k$ normalization function for $p^*(z|x^V)$ to be a valid conditional probability mass function. The other steps of the \textit{Bipartite} solver~\eqref{eq:wynerdca_pz} and \eqref{eq:wynerdca_complement} naturally follow by taking the corresponding expectations of the above solution.

\section{Convergence Proof of the VI Solver}\label{appendix:pf_thm_am_solver_conv}
Here we explain the idea for the convergence proof. In each update, we solve a convex sub-problem and hence the solver converges irrelevant to the initial starting point. Without loss of generality, consider the $k^{th}$ step-update for the first source $P^{k}(X_1|Z)$ with other variables $\{P^{k-1}(X_j|Z)\}_{j\neq 1}$ fixed. In this setup, each term of \eqref{eq:am_solver_bound} is a convex function of $P^{k}(X_1|Z)$ (convex, linear and convex respectively), but the joint distribution $P^k_\theta(X^V)=\sum_{z\in\mathcal{Z}}P(z)\prod_{i=1}^VP^{k}(X_i|z)$ involved in the last term is non-convex with respect to $\{P(X_i|Z)\}_{i=1}^V$ and hence, the convergence is guaranteed to a local stationary point.

\section{Convergence Proof of the Bipartite Solver}\label{appendix:pf_thm_wynerdca_conv}
We start with the following descent lemma that assures non-increasing objective function values between consecutive updates of the solutions obtained from the \textit{Bipartite} solver.
\begin{lemma}\label{lemma:descent_wynerdca}
Define the objective function value evaluated at $x$ as $\mathcal{L}(x):=f(x)-g(x)$. Suppose that $f,g$ are convex and differentiable, then the sequence $\{x^k\}_{k\in\mathbb{N}}$, obtained from \eqref{eq:wynerdca} results in a non-increasing sequence $\{\mathcal{L}(x^k)\}_{k\in\mathbb{N}}$.
\end{lemma}
\begin{IEEEproof}
    According to DCA \eqref{eq:dca_basic}, the first-order necessary condition is:
\begin{IEEEeqnarray}{rCl}
0=\nabla f(x^{k+1})-\nabla g(x^k),\IEEEeqnarraynumspace\IEEEyesnumber\label{eq:appendix_eq_fonc}
\end{IEEEeqnarray}
Then consider the following:
\begin{IEEEeqnarray}{rCl}
    \mathcal{L}(x^k)\nonumber&=&f(x^k)-g(x^k)\nonumber\\
    &\geq& f(x^{k+1}) +\langle\nabla f(x^{k+1}),x^{k+1}-x^k \rangle-g(x^k)\IEEEeqnarraynumspace\IEEEyesnumber\IEEEyessubnumber*\label{eq:appendix_pf_descent_fcvx}\\
    &=&f(x^{k+1})+ \langle \nabla g(x^k),x^{k+1}-x^k \rangle-g(x^k)\IEEEeqnarraynumspace\IEEEyessubnumber\label{eq:appendix_pf_descent_fonc}\\
    &\geq&f(x^{k+1}) -g(x^{k+1})=\mathcal{L}(x^{k+1}),\IEEEeqnarraynumspace\IEEEyessubnumber\label{eq:appendix_pf_descent_gcvx}
\end{IEEEeqnarray}
where \eqref{eq:appendix_pf_descent_fcvx} is due to the convexity of $f$, \eqref{eq:appendix_pf_descent_fonc} follows by the first order necessary condition \eqref{eq:appendix_eq_fonc}, and \eqref{eq:appendix_pf_descent_gcvx} is due to the convexity of $g$. The derivation shows that $\mathcal{L}(x^k)\geq \mathcal{L}(x^{k+1})$. Therefore, we conclude that $\{\mathcal{L}(x^k)\}_{k\in\mathbb{N}}$ is a non-increasing sequence.
\end{IEEEproof}
The convergence of the \textit{Bipartite} solver simply follows by Lemma \ref{lemma:descent_wynerdca} since the consecutive updates does not increase the objective function values, then there exists a $k_0>N_0\in\mathbb{N}$ sufficiently large such that for all $k'>k_0$, $P^{k'}_{z|x^V}$ converges to a stationary point $w^*\in\Omega^*:=\{w|\nabla \mathcal{L}(w)=0\}$. But since $\mathcal{L}(P_{z|x^V})$ is non-convex, the set of critical points $\Omega^*$ includes both local stationary points and the set of global minima.

\section{Exponential Family of Common Information}\label{appendix:exp_comm_info}
Here, we derive the exponential family of common information for the \textit{Bipartite} form only. As for the \textit{Variational} form, it can be considered a special case of the \textit{Bipartite} form with the configurations $\kappa_{[V]}$ and the additional conditional independence imposed as described in Section \ref{subsec:vi_unknown}.
Conditioned on a information source $X_S$, the common information in canonical form of the exponential family has the following log-likelihood:
\begin{IEEEeqnarray}{rCl}
    \log{P_\theta(Z|X_S)}=\log{h(Z)}+\boldsymbol{\eta}(\theta_{X_S})\cdot \boldsymbol{t}(Z)-A(\theta_{X_S}),\quad \forall S\in\Pi_V,
\end{IEEEeqnarray}
where $h(\cdot)$ is a normalization function; $\boldsymbol{\eta}$ the natural parameters of the member distribution, and $\boldsymbol{t}(z)$ the sufficient statistics; $A(\cdot)$ is known as the cumulant function; The operator $\cdot$ denotes an inner-product operator. 
Now, let us denote the natural parameters of a pre-determined reference density $P_\theta(Z)$ with $\theta_0$; whereas $\theta_{S}$ for the parameters of source $X_S$. Then the log-likelihood ratios can be expressed as:
\begin{IEEEeqnarray}{rCl}
    \log{\frac{P_\theta(Z|X_S)}{P_\theta(Z)}}=\left[\boldsymbol{\eta}(\theta_{S})-\boldsymbol{\eta}(\theta_{0})\right]\cdot \boldsymbol{t}(z) -\left[{A}(\theta_{S})-A(\theta_0)\right].
\end{IEEEeqnarray}
Then according to the projection \eqref{eq:unknown_bipartite_prob}, we have:
\begin{IEEEeqnarray}{rCl}
&&\log{P_\theta(Z|X^V)}\nonumber\\
&=&\log{h(Z)}+\boldsymbol{\eta}(\theta_0)\cdot \boldsymbol{t}(Z) -A(\theta_0)\nonumber\\
&&+\sum_{S\in\Pi_V}\kappa_S\left\{\left[\eta(\theta_S)+\boldsymbol{\eta}(\theta_{S^c})-2\boldsymbol{\eta}(\theta_0)\right]\cdot \boldsymbol{t}(Z)-\left[A(\theta_S)+A(\theta_{S^c})-2A(\theta_0)\right]\right\}\nonumber\\
&=&\log{h(Z)}+\left\{\boldsymbol{\eta}(\theta_0)+\sum_{S\in\Pi_V}\left[\boldsymbol{\eta}(\theta_S)+\boldsymbol{\eta}(\theta_{S^c})-2\boldsymbol{\eta}(\theta_0)\right]\right\}\cdot \boldsymbol{t}(Z)-\left[A(\theta_S)+A(\theta_{S^c})-2A(\theta_0)\right],
\end{IEEEeqnarray}
By defining the equivalent natural parameter as:
\begin{IEEEeqnarray}{rCl}
    \boldsymbol{\eta}_{eq}:=\boldsymbol{\eta}(\theta_0)+\sum_{S\in\Pi_V}\left[\boldsymbol{\eta}(\theta_S)+\boldsymbol{\eta}(\theta_{S^c})-2\boldsymbol{\eta}(\theta_0)\right],
\end{IEEEeqnarray}
the log-likelihood (with proper normalization) distributed as the same member distribution as the reference prior. For example, in Gaussian model, the natural parameter is $\boldsymbol{\eta}(\theta)=\begin{bmatrix}\boldsymbol{\Sigma}^{-1}\boldsymbol{\mu}&\boldsymbol{\Sigma}^{-1}\end{bmatrix}^T$, then for $P_\theta(Z)=\mathcal{N}(\boldsymbol{0},\boldsymbol{I})$, we have:
\begin{IEEEeqnarray}{rCl}
\boldsymbol{\Sigma}^{-1}_{eq}\boldsymbol{\mu}_{eq}&=&\sum_{S\in\Pi_V}\kappa_S\left(\boldsymbol{\Sigma}^{-1}_S\boldsymbol{\mu}_S+\boldsymbol{\Sigma}^{-1}_{S^c}\boldsymbol{\mu}_{S^c}\right)\nonumber\\
\boldsymbol{\Sigma}^{-1}_{eq}&=&\boldsymbol{I}+\sum_{S\in\Pi_V}\kappa_S\left[\boldsymbol{\Sigma}_S^{-1}+\boldsymbol{\Sigma}_{S^c}^{-1}-2\boldsymbol{I}\right],\nonumber
\end{IEEEeqnarray}
and we obtain the equivalent mean as $\boldsymbol{\mu}_{eq}:=\left[\boldsymbol{I}+\sum_{S\in\Pi_V}\kappa_S\left(\boldsymbol{\Sigma}^{-1}_S+\boldsymbol{\Sigma}_{S^c}^{-1}-2\boldsymbol{I}\right)\right]^{-1}\left(\sum_{\pi\in\Pi_V}\boldsymbol{\Sigma}_{\pi}^{-1}\boldsymbol{\mu}_{\pi}+\boldsymbol{\Sigma}_{\pi^c}^{-1}\boldsymbol{\mu}_{\pi^c}\right)$.

\section{Correlation Optimization for Multi-Modal Data}\label{sec:appendix_mi_min}
The proof of the lower bound of mutual information in contrastive learning is based on the argument of a discriminative ``critic'', whose responsibility is to identify the optimal ``positive'' sample from other $N-1$ negative ones, followed by an application of the weak-law of large number~\cite{tian2020contrastive,xu2022MFLVC,oord2018representation,poole2019variational}. Here, we present an alternative proof.

Without loss of generality, consider a batch of $N$ sample pairs $S_+=\{w^{(n)}_1,w^{(n)}_2\}_{n=1}^N$, assume that $T_N(w_i^{(j)})=\frac{1}{N},\forall i\in\{1,2\},\forall j\in[N]$, then the empirical joint distribution is given by $T_N(W_1,W_2)=\frac{1}{N}\boldsymbol{1}\{i=j|(w_1^{(i)},w_2^{(j)}),\forall i,j\in[N]\}$, and we have:
\begin{IEEEeqnarray}{rCl}
    I_{T_N}(W_1,W_2)&=&\mathbb{E}_{T_N(W_1,W_2)}\left[\log{\frac{T_N(W_2|W_1)}{T_N(W_2)}}\right]\nonumber\\
    &=&\mathbb{E}_{T_N(W_1,W_2)}\left[\log{\frac{T_N(W_2|W_1)}{Q_N(W_2|W_1)}\frac{Q_N(W_2|W_1)}{T_N(W_2)}}\right]\nonumber\\
    &=&\mathbb{E}_{T_N(W_1)}\left\{D_{KL}\left[T_N(W_2|W_1)\parallel Q_N(W_2|W_1)\right]\right\}-\mathbb{E}_{T_N(W_1)}\left\{D_{KL}[T_N(W_2)\parallel Q_N(W_2|W_1)]\right\}\nonumber\\
    &\leq&\mathbb{E}_{T_N(W_1)}\left\{D_{KL}\left[T_N(W_2|W_1)\parallel Q_N(W_2|W_1)\right]\right\}\nonumber\\
    &=&-\mathbb{E}_{T_N(W_1)}\left[\boldsymbol{1}\{i=j\}\log{Q_N(W_2^{j}|W_1^{i})}\right],\IEEEeqnarraynumspace\IEEEyesnumber\label{eq:appendix_our_derivation_contrast}
\end{IEEEeqnarray}
where the inequality is due to the non-negativity of the KL divergence, and $Q_N$ is a parameterized distribution. We conclude that to estimate the mutual information, one can minimize the upper bound. This is equivalent to solving the maximization problem:
\begin{IEEEeqnarray}{rCl}
    Q^*_N:=\underset{\{h_+,h_-\}\in\mathcal{H}}{\arg\max}\, \frac{1}{N}\sum_{n=1}^N\log{\frac{{h_{+}(w_1^{(n)},w_2^{(n)})}}{{h_+(w_1^{(n)},w_2^{(n)})}+\sum_{k\neq n}^{N}{h_{-}(w_1^{(n)},w_2^{(k)})}}}.
\end{IEEEeqnarray}
That is, the parameterized distribution should output the highest similarity for the paired sample, and with low similarity to all other pairs. A popular implementation of the above in recent literature is parameterizing $Q_N$ as a neural network whose output activation function is softmax with $N$ categories. This means that $Q_N$ is parameterized as a categorical distribution. Then the network can be optimized by imposing a categorical cross entropy loss function with the class labels corresponding to the indices of the paired sample. 


\ifCLASSOPTIONcaptionsoff
  \newpage
\fi

\end{document}